\documentclass[12pt,letterpaper]{article}
\usepackage{jheppub}

\usepackage{graphicx}
\usepackage{bbm}
\usepackage{amsmath}
\usepackage{amssymb}
\usepackage{mathtools}

\usepackage[labelformat=simple]{subcaption}

\usepackage{ifpdf}
\newcommand{\alt}[2]{\texorpdfstring{#1}{#2}}

\allowdisplaybreaks

\newcommand{\noeq}{\mathrel{\phantom{=}}}

\newcommand{\Fself}{\hat{F}_\text{self}}
\newcommand{\Lhd}{\mathcal{L}_\text{hd}}
\newcommand{\Shd}{S_\text{hd}}
\newcommand{\Thd}{T_{\!\text{hd}}}
\newcommand{\THD}{T^{\!\text{hd}}}
\newcommand{\ap}{\alpha'}

\newcommand{\secref}[1]{\S\ref{#1}}

\def\be{\begin{equation}}
\def\ee{\end{equation}}
\def\ba{\begin{align}}
\def\ea{\end{align}}

\begin{document}
\begin{titlepage}

\setcounter{page}{1} \baselineskip=15.5pt \thispagestyle{empty}

{\flushright ACFI-T22-09 \\ }

\bigskip\

\vspace{2cm}
\begin{center}
{\LARGE \bfseries Derivative Corrections to \\ \vspace{0.3cm} Extremal Black Holes with Moduli}

 \end{center}
\vspace{1cm}

\begin{center}
\scalebox{0.95}[0.95]{{\fontsize{14}{30}\selectfont Muldrow Etheredge and Ben Heidenreich}}
\end{center}

\begin{center}
\vspace{0.25 cm}
\textsl{Amherst Center for Fundamental Interactions\\Department of Physics, University of Massachusetts, Amherst, MA 01003 USA}\\

\vspace{0.25cm}
\end{center}

\vspace{1cm}
\noindent

We derive formulas for the leading mass, entropy, and long-range self-force corrections to extremal black holes due to higher-derivative operators. These formulas hold for black holes with arbitrary couplings to gauge fields and moduli, provided that the leading-order solutions are static, spherically-symmetric, extremal, and have nonzero horizon area. To use these formulas, both the leading-order black hole solution and the higher-derivative effective action must be known, but there is no need to solve the derivative-corrected equations of motion. We demonstrate that the mass, entropy and self-force corrections involve linearly-independent combinations of the higher-derivative couplings at any given point in the moduli space, and comment on their relations to various swampland conjectures.

\vspace{1.1cm}

\bigskip
\noindent\today

\end{titlepage}

\setcounter{tocdepth}{2}

\hrule
\tableofcontents

\bigskip\medskip
\hrule
\bigskip\bigskip
\section{Introduction}
The study of the Weak Gravity Conjecture (WGC) \cite{Arkani-Hamed:2006emk,Harlow:2022gzl}---a prototype for the broader Swampland Program \cite{Vafa:2005ui,Palti:2019pca}---has led to a number of educated guesses about the structure of quantum gravity that have so far been successfully tested in many examples.

One such prediction is that higher-derivative corrections to the low-energy effective action \emph{decrease} the mass of extremal black holes at fixed charge~\cite{Kats:2006xp}.\footnote{Extremal black holes are here defined as the lightest static, spherically-symmetric, black holes of given electric and magnetic charges. All black holes in this paper are assumed to be static and spherically-symmetric.}
This can be
motivated in part by the sublattice and tower versions of the Weak Gravity Conjecture (WGC) \cite{Heidenreich:2016aqi,Andriolo:2018lvp}, which require a tower of states of arbitrarily large charge, each with charge-to-mass ratio at least as large as that of a parametrically heavy extremal black hole.
 In perturbative string theory, the lightest states are well described by string oscillation modes, and thus the tools of perturbative string theory can be used to gather evidence for the WGC~\cite{Heidenreich:2016aqi,BenMatteoProof}. However, as sufficiently excited strings collapse into black holes, the heavier states required by these conjectures cannot be probed in the same way. Instead, the conjectures then hinge on the spectrum of charged black holes, and in particular whether higher-derivative corrections make them lighter (or, at least, not heavier) at fixed charge.

The Repulsive Force Conjecture (RFC)~\cite{Palti:2017elp,Lust:2017wrl,Lee:2018spm,Heidenreich:2019zkl} is a close relative of the WGC that requires the existence self-repulsive states, i.e., states that exert a repulsive or vanishing long-range force on their identical copy (called a ``self-force") when separated from it by a parametrically large distance. As with the WGC, the light string excitations satisfy a stronger sublattice/tower version of the RFC~\cite{Heidenreich:2019zkl,BenMatteoProof}, but as before highly excited strings collapse into black holes, hence for heavier states the conjectures hinge on the self-forces of charged black holes. In particular, at the two-derivative level static, spherically-symmetric extremal black holes have the remarkable property that their self-force vanishes (see, e.g.,~\cite{Heidenreich:2020upe}), hence higher-derivative corrections must make them self-repulsive (or, at least, not self-attractive) to satisfy the sublattice/tower RFC.

Though less directly connected to an existing swampland conjecture, it has also been suggested~\cite{Cheung:2018cwt} that higher-derivative corrections to the black hole \emph{entropy} should be non-negative, heuristically because derivative corrections represent the effects of heavy modes and the possibility of exciting these modes leads to a larger number of microstates.

The effect of higher-derivative corrections on extremal black holes has been studied most thoroughly in the absence of moduli (i.e., scalar fields with vanishing potential), where the WGC and RFC become the same. The four-derivative corrections to electrically-charged Reissner-Nordstr\"om black holes were obtained in \cite{Kats:2006xp}, and generalized to solutions with arbitrarily many gauge fields in~\cite{Jones:2019nev}. Similar calculations were done for Kerr black holes in \cite{Reall:2019sah,Aalsma:2020duv,Aalsma:2021qga}, and for black holes in non-asymptotically-flat backgrounds such as AdS in \cite{Aalsma:2021qga}. Discussions and calculations of entropy corrections can be found in \cite{Cvetic:1995bj,Sahoo:2006pm,Cheung:2018cwt,Loges:2019jzs,Cremonini:2019wdk}.

On the other hand, well-understood string compactifications do have moduli. Previous works have largely focused on the case where a single dilaton modulus is present, beginning with \cite{Natsuume:1994hd}, where the extremal mass corrections were calculated in heterotic string theory, and \cite{Kats:2006xp}, where these corrections were shown to decrease the mass and generate a repulsive self-force on extremal black holes. A more general bottom-up analysis of dilatonic couplings was carried out in~\cite{Loges:2019jzs}, where four-dimensional dyonic black holes were also considered.

Unfortunately, electrically-charged extremal black holes coupled to a dilaton modulus are ``small'', in that the corresponding solution to the two-derivative effective action has a horizon of vanishing surface area. What actually occurs near the horizon of a small black hole depends on curvature corrections that are arbitrarily high order, hence it is a UV-sensitive question that cannot be answered using the low-energy effective action alone, even when supplemented by its low-order derivative corrections.\footnote{Similar points were made in, e.g., \cite{Cvetic:1995bj}. Note that the analysis of~\cite{Natsuume:1994hd}, referenced in~\cite{Kats:2006xp}, is at string-tree-level, where derivative corrections are fortuitously insensitive to the singular horizon. This insensitivity does not, however, extend to string-loop corrections since the dilaton is infinite on the horizon, so the derivative expansion is not under control, see~\S\ref{sec.EMdexample} for further discussion.\label{fn:dilatonCorrections}}
 For this reason, in this paper we focus on corrections to ``large" black holes---black holes with horizons of nonvanishing area.

Extremal black holes with dilatonic couplings can be ``large" when they are dyonically charged. However, dyonic charge only exists in four dimensions, whereas to our knowledge complete string-theory-derived four-derivative effective actions have only been computed in high dimensions, see for example \cite{Gross:1986mw} for the ten-dimensional heterotic case. While these couplings can be dimensionally reduced as in, e.g., \cite{Cremonini:2021upd,Cano:2021nzo,Ortin:2021win}, there are potentially important subtleties in this procedure. For instance, as discussed in \cite{Green:1997di,Green:1997as}, Kaluza-Klein (KK) reducing eleven-dimensional M-theory on a circle or torus and integrating out the massive KK modes at one loop generates further derivative corrections beyond those present in the eleven-dimensional effective action. Thus, it is not sufficient to just dimensionally reduce the four-derivative terms in the ten-dimensional effective action---one must also integrate out the KK-modes, or at least argue that their contributions are less important than the dimensionally-reduced derivative corretions. To our knowledge, this has yet to be done in the literature, nor have the corrections been obtained directly from the compactified worldsheet sigma model.

\bigskip

In the absence of this crucial string theory input, in this paper we focus on the problem of determining the mass, self-force, and entropy corrections to large extremal black holes once the four-derivative effective action is known. In this context, we are able to provide a very general answer, assuming only that the solution is static and spherically-symmetric. The formulas we obtain hold for black holes with arbitrarily many gauge fields and moduli, arbitrary four-derivative operators, and arbitrary couplings between the moduli and gauge fields, as long as the extremal two-derivative solutions have horizons with non-vanishing surface area. Using these formulas requires only the original two-derivative solution along with the four-derivative effective action and its functional first-derivatives (such as the stress tensor). In particular, the derivative corrected solution is not needed. The mass and force corrections can be expressed even more simply, depending only on the four-derivative Lagrangian density evaluated on the two-derivative solution.

As a preview, by a direct attack on the equations of motion, we find the following explicit formulas for the corrections to the mass, entropy and self-force of an extremal black hole of fixed charge:
\begin{subequations} \label{eqn:basicResults}
\begin{align}
\delta M &= - \ap V_{d-2}  \int_{r_h}^\infty \biggl(\Thd{}^t_t + F_{t r}^A \frac{\delta \Shd}{\delta F^A_{tr}} \biggr) \mathcal{R}^{d-2} \sqrt{|g_{t t} g_{r r}|} dr ,  \label{eq.formulaM} \\
\delta \mathcal{S} &= 2\pi \ap V_{d-2}  \!\Biggl[-\frac{\mathcal{R}^d}{(d-3)^2} \biggl( \Thd{}^t_t+F^A_{tr} \frac{\delta \Shd}{\delta F^A_{tr}}\biggr) +\mathcal{R}^{d-2} \frac{\delta \Shd}{\delta R^{tr}\!_{tr}}\Biggr]_{r=r_h},\label{eq.formulaS}\\
\delta \Fself &= -2\ap V^2_{d-2}  \int_{r_h}^\infty  \Bigl((d-2)\Thd{}^r_r+\Thd{}^i_i \Bigr) \mathcal{R}^{2d-5} |g_{t t}| \sqrt{g_{r r}} dr . \label{eq.formulaF}
\end{align}
\end{subequations}
Here $\alpha'$ is a formal derivative-expansion parameter in the action $S = S_2 + \alpha' \Shd$, $\Thd$ is the stress tensor associated to $\Shd$ where $\Thd{}^i_i = \sum_{i=1}^{d-2} \Thd{}_i^i$ denotes the partial trace over angular directions, $V_{d-2}$ is the unit $(d-2)$-sphere volume, and $\Fself$ is the rationalized coefficient of the long-range self-force, $\mathbf{F}_\text{self}(r) = \frac{\Fself}{V_{d-2}} \frac{\mathbf{\hat{r}}}{r^{d-2}}+ \ldots$. The functional derivatives with respect to $F_{\mu\nu}^A$ and $R_{\mu\nu\rho\sigma}$ are normalized as
\be
\delta \Shd = \int d^d x \sqrt{-g}\, \biggl[\frac{1}{2} \frac{\delta \Shd}{\delta F^A_{\mu \nu}} \delta F^A_{\mu \nu} + \frac{1}{4} \frac{\delta S}{\delta R_{\mu \nu \rho \sigma}} \delta R_{\mu \nu \rho \sigma} \biggr] \,,
\ee
each one having the same symmetries as the tensor in question. The corrections are to be evaluated by substituting the uncorrected extremal black hole solution, with the spherically-symmetric metric
\begin{equation}
d s^2 = g_{t t} d t^2 + g_{r r} dr^2 +\mathcal{R}(r)^2 d\Omega^2 , \label{eq.ansatzfromintro}
\end{equation}
into \eqref{eq.formulaF}--\eqref{eq.formulaS} and evaluating the radial integral (in the mass and force cases) or taking the near-horizon limit $r\to r_h$ (in the entropy case). 

In fact, the mass and entropy corrections can be more simply expressed in terms of the higher-derivative Lagrangian density $\Lhd$ itself (with $\Shd = \int d^d x \sqrt{-g} \Lhd$)\footnote{Unlike~\eqref{eq.formulaM}, we have only proven \eqref{eqn:simpMassFormula} under simplifying assumptions that are valid up to four-derivative order, see~\S\ref{subsec:simplifications} for details, but we strongly suspect that it holds in general.}
\begin{subequations} \label{eqn:simpFormulas}
\begin{align}
  \delta M &= - \alpha' V_{d - 2} \int_{r_h}^{\infty} \Lhd
  \mathcal{R}^{d - 2} \sqrt{| g_{t t} g_{r r} |} d r , \label{eqn:simpMassFormula} \\
  \delta \mathcal{S} &= \left. - \frac{2 \pi \alpha'}{(d - 3)^2} V_{d - 2}
  \mathcal{R}^d \Lhd \right|_{r = r_h} . \label{eqn:simpEntropyFormula} 
\end{align}
\end{subequations}
These formulas---which we arrive at indirectly---are so simple and elegant that there is very likely some general principle underlying them, but we leave this interesting question to future work.

Note that both of our mass formulas were previously derived in the absence of moduli but generalized to 4d rotating black holes, see~\cite{Aalsma:2020duv} in the case of~\eqref{eq.formulaM} and~\cite{Cheung:2019cwi,Arkani-Hamed:2021ajd} in the case of \eqref{eqn:simpMassFormula}. We know of no previous work on these formulas in the presence of moduli. Likewise, the entropy and force formulas \eqref{eq.formulaS}, \eqref{eq.formulaF}, and \eqref{eqn:simpEntropyFormula}  are completely new results to our knowledge.

Using these formulas, we show that the extremal force, mass, and entropy corrections depend on the four-derivative operators in independent ways, and it is possible to have the mass, self-force, and entropy corrections all take on arbitrary signs relative to each other. This agrees with some previous results in the literature. For example, in \cite{Cremonini:2021upd} it was shown that extremal mass and extremal force corrections can take different signs. However, it seems naively in tension with the results of \cite{Cheung:2018cwt,Goon:2019faz}, where it was shown that the entropy correction at fixed mass and charge is positive near extremality if and only if the extremal mass correction is negative. The resolution is that the extremal entropy correction \eqref{eq.formulaS}, \eqref{eqn:simpEntropyFormula} is not the same as the entropy correction at fixed mass and charge near extremality, as previously argued in~\cite{McPeak:2021tvu}. Indeed, the latter generally diverges whereas the former is finite. Per the general results of~\cite{Goon:2019faz} (which we reproduce here), the divergent portion is fixed by the extremal mass correction. Thus, the positivity (or not) of the extremal entropy correction \eqref{eq.formulaS}, \eqref{eqn:simpEntropyFormula} remains an interesting and relatively unexplored question, whereas the positivity of the entropy correction at fixed mass and charge near extremality is completely equivalent to the negativity of the extremal mass correction.

Our paper is structured as follows. In \secref{sec.corrections}, we derive the force, mass, and entropy correction formulas for static, spherically-symmetric extremal black holes. In \secref{sec.independence}, we show that the extremal mass and entropy corrections are a priori independent and explain how this can be consistent with the general result of \cite{Goon:2019faz} relating the extremal mass correction to the entropy correction at fixed mass and charge near extremality. In \secref{sec.examples} we illustrate our method by examining a few specific examples and comparing with existing results in the literature.  We conclude by highlighting a few interesting directions for future research in~\secref{sec.conclusions}. In appendix \ref{sec.basis} we derive a minimal basis of independent four-derivative operators in the presence of moduli and arbitrarily many gauge fields. Appendices~\ref{app:Riemann} and~\ref{app:stresstensor} contain a few formulas that are helpful for computing the corrections in specific examples.

\section{Self-force, mass, and entropy corrections \label{sec.corrections}}
We now compute the leading derivative corrections to the self-force, mass, and entropy of non-rotating extremal black holes. We assume that the black holes in question are static and spherically-symmetric---as in familiar examples of non-rotating, extremal black hole solutions\footnote{We know of no theorem that non-rotating extremal black hole solutions must be static and spherically symmetry at the two-derivative level, much less accounting for derivative corrections, so it would be interesting---if technically very difficult---to relax this assumption.}---and that the cosmological constant vanishes. For simplicity, we also initially assume that the black holes carry only electric charge, even though magnetic and dyonic charges are also possible in four-dimensions. As we argue later, our final results generalize without any modifications to the dyonic/magnetic case.

\subsection{The low-energy effective action\label{sec.eactionansatz}}

Since we are interested in static, spherically-symmetric, electrically-charged black hole solutions, at the two-derivative level we can restrict our attention to an effective action of the form
\begin{align}
S&=\int d^dx\sqrt{-g} \mathcal L_2, &
\mathcal L_2&=\frac 1{2\kappa^2}R-\frac 12 G_{ab}(\phi)\nabla \phi^a\cdot\nabla \phi^b-\frac 12 f_{AB}(\phi)F^A\cdot F^B, \label{eq.twoderivaction}
\end{align}
as argued in~\cite{Heidenreich:2020upe}, where $p$-form dot product is defined as $G_p\cdot H_p\equiv \frac 1{p!}G^{\mu_1\dots\mu_p}H_{\mu_1\dots\mu_p}$, $a=1,\ldots,n_\phi$ label the moduli, $A=1,\ldots,n_A$ label a Cartan subalgebra of the gauge algebra and $G_{ab}(\phi)$ and $f_{AB}(\phi)$ are, respectively, the metric on moduli space and the (moduli-dependent) gauge kinetic matrix. Note that~\eqref{eq.twoderivaction} omits all charged and/or fermionic fields---which can be consistently truncated---as well as all massive fields---which have been integrated out.\footnote{Moreover, we assume that all massless neutral scalar fields are moduli, hence $V(\phi) = 0$. A massless, neutral scalar field with a non-vanishing potential would have a similar effect on black hole solutions to the derivative corrections that we study, but requires a separate analysis.} Also absent are fields and couplings that have no effect on static spherically-symmetric black hole backgrounds, such as higher-form gauge fields and their Chern-Simons couplings.

We now consider higher-derivative corrections to~\eqref{eq.twoderivaction}:
\begin{equation}
\mathcal{L} = \mathcal{L}_2 + \ap \Lhd + \ldots ,
\end{equation}
where $\Lhd$ contains the leading higher-derivative corrections and
$\ap$ is a formal order-counting parameter of negative mass dimension---notationally inspired but not necessarily related to $\ap$ in string theory. $\Lhd$ encodes the infrared consequences of a wide variety of UV physics, such as massive particles, extra dimensions, stringy physics, etc. The particular nature of this UV physics will not matter for our analysis, except that in the case of massive particles we assume that none of them become massless in a part of the moduli space visited by the black hole solution in question; otherwise, the extra massless particles must be incorporated into the action to maintain control of the effective field theory, an extra step that is beyond the scope of this paper.

Typically $\Lhd$ consists of four-derivative operators, but our general formulae will not depend on this. For illustration, as shown in equation \eqref{eq.Lleq4basis} of Appendix \secref{sec.basis}, the four-derivative operators that correct the mass and self-force of static spherically-symmetric electrically-charged black hole solutions can be put into the following form:
\begin{multline}
	\Lhd =a_{ABCD}(\phi)(F^A\cdot F^B)(F^C\cdot F^D)+\frac 14a_{AB}(\phi)F^A_{\mu \nu} F^B_{\rho\sigma}R^{\mu \nu \rho \sigma}+a(\phi)R_\text{GB}\\
+a_{abcd}(\phi)(\nabla \phi^a\cdot\nabla \phi^b)(\nabla\phi^c\cdot\nabla\phi^d)+a_{ABab}(\phi)(\nabla \phi^a\cdot\nabla\phi^b)(F^A\cdot F^B), \label{eq.LhdBasis}
\end{multline}
up to total derivatives, field redefinitions, and combinations of operators that have no effect on static spherically-symmetric electrically-charged black hole solutions, where $R_\text{GB}\equiv  R_{\mu \nu \rho \sigma}R^{\mu \nu \rho \sigma}-4R_{\mu \nu}R^{\mu \nu}+R^2$ is the Gauss-Bonnet term and $a_{A B C D}(\phi)$, $a_{A B}(\phi)$, etc., are \emph{a priori} general functions of the moduli. Thus, the set of effective operators relevant to problem at hand is both rich and enumerable; however, our results will not depend on $\Lhd$ taking the form~\eqref{eq.LhdBasis}.

Our analysis of the effects of these operators will be semiclassical. As argued in~\cite{Charles:2019qqt,Arkani-Hamed:2021ajd}, one-loop effects can be important or even dominant in the four-dimensional case, so our results must be treated with caution in $d=4$.

\subsection{Black hole ansatz and equations of motion\label{sec.eeom}}

A general static spherically-symmetric electrically-charged black hole solution takes the form
\begin{equation}
\begin{aligned}
ds^2&=-e^{2\psi(r)}f(r)dt^2+e^{-\frac 2{d-3}\psi(r)}\left[\frac{dr^2}{f(r)}+r^2 d\Omega^2_{d-2}\right], \\
F^A & =F^A_{tr}(r) dt\wedge dr, \\
\phi^a &= \phi^a(r) ,
\end{aligned} \label{eq.ansatz}
\end{equation}
where $d\Omega^2_{d-2}$ is the round metric of unit radius on the transverse $S^{d-2}$ and we choose the same gauge as in, e.g.,~\cite{Heidenreich:2020upe}, without yet making use of the equations of motion.

The equations of motion for the gauge-fields read
\begin{align}
d\star \mathcal F_A&=0,&d F^A&=0, &
\text{where}\quad
\mathcal F_A&=f_{AB}F^B- \ap \frac{\delta \Shd}{\delta F^A} ,\label{eq.gaugeeom}
\end{align}
where $\frac{\delta \Shd}{\delta F^A}$ is the covariant functional derivative of $\Shd = \int d^d x \sqrt{-g} \Lhd$, defined via the functional variation
\be
\delta \Shd = \int d^d x \sqrt{-g}\, \delta F^A \cdot \frac{\delta \Shd}{\delta F^A} = \int d^d x \sqrt{-g}\, \frac{1}{2} \delta F^A_{\mu \nu} \frac{\delta \Shd}{\delta F^A_{\mu \nu}},
\ee
and we assume that $\Lhd$ depends only on the field strength $F^A$, not directly on the gauge potential $A^A$.\footnote{In particular, this excludes higher-derivative Chern-Simons terms, see appendix~\ref{sec.basis} for a justification for this omission.}

Using spherical symmetry, the $t r$ component of $\mathcal{F}_A$ is completely fixed by $\psi$ and the electric charge of the black hole $Q_A$:
\begin{align}
Q_A &= \oint_{S^{d-2}}\star \mathcal F_A, &\Longrightarrow \qquad \mathcal F_{A t r}=-\frac{Q_Ae^{2\psi}}{V_{d-2}r^{d-2}} ,
\end{align}
where $V_{d-2} = \frac{2 \pi^{\frac{d-1}{2}}}{\Gamma(\frac{d-1}{2})}$ is the area of a unit-radius $S^{d-2}$ sphere.
Thus, we obtain\footnote{Note that while $F^A \propto d t \wedge d r$ is required by spherical symmetry in $d>4$, in 4d spherical symmetry also permits an angular $\sin \theta d \theta \wedge d \phi$ component. However, this vanishes when the magnetic charge $Q^A _m = \oint_{S^{d-2}} F_2$ vanishes, as assumed in this section.}
\be
F^A = f^{AB} \biggl(-\frac{Q_Be^{2\psi}}{V_{d-2}r^{d-2}} +\ap \frac{\delta \Shd}{\delta F^{B t r}} \biggr) dt\wedge dr ,\label{eq.F}
\ee
where $f^{AB}(\phi)$ denotes the inverse of the gauge kinetic matrix $f_{AB}(\phi)$.

The moduli equations of motion and Einstein equations are
\be
\begin{aligned}
\nabla^2\phi^a+\Gamma^a_{bc}(\partial \phi^b\cdot\partial\phi^c) &= G^{ab} \Bigl(\frac{1}{2} f_{AB,b}F^A\cdot F^B- \ap  \frac{\delta \Shd}{\delta \phi^b}\Bigr),\\
R_{\mu \nu}-\frac 12g_{\mu \nu}R = \kappa^2 T_{\mu \nu} &= \kappa^2\left(G_{ab}\partial_\mu \phi^a\circ\partial_\nu \phi^b+ f_{AB}F^A_\mu \circ F^B_\nu+ \ap \THD_{\mu \nu}\right),
\end{aligned}
\ee
where $G^{a b}(\phi)$ denotes the inverse of the metric on moduli space $G_{a b}(\phi)$ and\footnote{As before, functional derivatives are defined covariantly, so that $\delta \Shd = \int d^d x \sqrt{-g} \frac{\delta \Shd}{\delta g^{\mu \nu}} \delta g^{\mu\nu}+\int d^d x \sqrt{-g} \frac{\delta \Shd}{\delta \phi^a} \delta \phi^a$. In general, writing the functional derivatives of $\Shd$ in terms of ordinary derivatives of $\Lhd$ requires integration by parts, e.g., $\frac{\delta \Shd}{\delta \phi^a} = \frac{\partial \mathcal \Lhd}{\partial \phi^a}-\nabla_\mu\frac{\partial \mathcal \Lhd}{\partial( \nabla_\mu \phi^a)}$ in the case where $\Lhd$ contains no second derivatives of $\phi^a$.}
\begin{align}
\begin{aligned}
\Gamma^a_{bc} &\equiv  \frac 12 G^{ad}(G_{bd,c}+G_{cd,b}-G_{bc,d}), &
\THD_{\mu\nu} &\equiv  -2 \frac{\delta \Shd}{\delta g^{\mu \nu}} , \\
\omega_\mu\circ\chi_\nu &\equiv  \omega_\mu\cdot \chi_\nu-\frac 12 g_{\mu \nu}\omega\cdot \chi, &
\omega_\mu\cdot\chi_\nu &\equiv  \frac 1{p!}\omega_{\mu \nu_1\dots\nu_p}\chi_\nu\!^{\nu_1\dots\nu_p}, 
\end{aligned}
\end{align}
for arbitrary $(p+1)$-forms $\omega$ and $\chi$.

Applying the ansatz~\eqref{eq.ansatz} and computing the associated Ricci tensor, we obtain:
\begin{subequations}
\begin{align}
\frac{1}{r^{d-2}}(r^{d-2}f\phi'^a)'+f\Gamma^a_{bc}\phi'^b\phi'^c &= e^{-\frac{2\psi}{d-3}} G^{ab} \biggl(\frac{1}{2} f_{AB,b}F^A\cdot F^B-\ap \frac{\delta \Shd}{\delta \phi^b} \biggr),\\
\frac{1}{r^{2d-5}}\left(r^{d-2}\left[r^{d-3}(1-f)\right]'\right)' &= -2\kappa^2 e^{-\frac{2\psi}{d-3}} \left(T^r_r+\frac{1}{d-2}T^i_i\right),\\
\frac{1}{r^{d-2}}\Bigl[r^{d-2} f\psi'+(d-3) r^{d-3}(1-f)\Bigr]' &= -\frac{d-3}{d-2}\kappa^2 e^{-\frac{2\psi}{d-3}} (T^t_t+T^r_r),\\
\psi'\left[f\psi'+f'\right]+\frac{(d-3)^2}{r^2}(1-f)-\frac{d-3}rf' &=-2 \frac{d-3}{d-2} \kappa^2 e^{-\frac{2\psi}{d-3}}  T^r_r,
\end{align}
\end{subequations}
where primes denote $r$-derivatives, $T=T^\mu_\mu$, and $T^i_i = \sum_{i=1}^{d-2} T_i^i$ denotes the partial trace of $T^\mu_\nu$ over the angular directions.

To simplify these equations, it is convenient to define the inverse radial variable
\begin{align}
z\equiv &\;\frac 1{(d-3)V_{d-2} r^{d-3}}&\Rightarrow&&dz&=-\frac 1{V_{d-2} r^{d-2}}dr,
\end{align}
as well as the function
\begin{align}
\chi(z)\equiv &\;\frac{1-f}z&\Leftrightarrow&&f&=1-z\chi(z).
\end{align}
In terms of these, the equations of motion become
\begin{subequations}\label{eq.zeom}
\begin{align}
\frac d{dz}(f\dot\phi^a)+f\Gamma^a_{bc}\dot\phi^b\dot\phi^c &= e^{2 \psi} A^2\, G^{ab} \left(\frac 12f_{AB,b}F^A\cdot F^B-\ap \frac{\delta \Shd}{\delta \phi^b} \right), \label{eq.phieqn} \\
\ddot\chi &= -\frac{2k_N e^{2 \psi} A^2}{(d-3)z} \bigl((d-2)T^r_r+T^i_i\bigr), \label{eq.chieqn} \\
\frac d{dz}\bigl(f\dot\psi-\chi\bigr) &=  -k_N e^{2 \psi} A^2 \bigl(T^t_t+T^r_r\bigr), \label{eq.psieqn} \\
\dot\psi\left[f\dot \psi+\dot f\right]-\dot \chi &= - 2 k_N e^{2 \psi} A^2\, T^r_r, \label{eq.conseqn}
\end{align}
\end{subequations}
where dots denote $z$-derivatives,
\begin{equation}
	A(z) \equiv  V_{d-2} r^{d-2} e^{-\frac{d-2}{d-3} \psi} \label{eq.Aofz}
\end{equation}
is the $z$-dependent area of the $S^{d-2}$, and
\begin{equation}
	k_N \equiv \frac{d-3}{d-2} \kappa^2  
\end{equation}
is the rationalized Newtonian force constant. Note that \eqref{eq.conseqn} is a constraint equation at leading order in the derivative expansion: differentiating it gives a linear combination of the other equations.

To study the event horizon, we rewrite the metric in infalling coordinates:
\be
d s^2 = - \frac{F(\rho) d v^2}{\mathcal{R}(\rho)^{2(d-3)}}  + \frac{2 d v d \rho}{(d-3) \mathcal{R}(\rho)^{d-4}} + \mathcal{R}(\rho)^2 d \Omega_{d-2}^2,\label{eq.infall}
\ee
where $\rho \equiv  r^{d-3}$, $\mathcal{R}(\rho) \equiv  r e^{-\frac{\psi}{d-3}}$, and $F(\rho) \equiv  r^{2(d-3)} f(r)$. Thus, a smooth horizon requires $F(\rho) \to 0$ at finite $\rho = \rho_h$ with $\mathcal{R}(\rho)$ remaining finite and non-zero. There is a residual gauge symmetry shifting $\rho$ by a constant while holding the form of $\mathcal{R}(\rho)$ and $F(\rho)$ fixed, so the value of $\rho_h$ is thus far meaningless.
By contrast, $F'(\rho_h) \ge 0$ is a gauge-invariant characteristic of the horizon. In particular, the product of the surface gravity $g_h$ times the horizon area $A_h$ is readily found to be $g_h A_h = \frac{d-3}{2} V_{d-2} F'(\rho_h)$, so $F'(\rho_h) = 0$ is the (quasi)extremal case in the terminology of~\cite{Heidenreich:2020upe}, whereas $F'(\rho_h) > 0$ is the subextremal case.

While in principle we can proceed in any gauge, it will be very convenient to make the gauge choice $\rho_h = F'(\rho_h)$, so that $\rho_h \ge 0$ with $\rho_h \to 0$ in the (quasi)extremal limit. In terms of $r$ and $f(r)$, this becomes\footnote{In the quasiextremal case we obtain the boundary condition $f(r=0) = \text{finite}$ instead.}
\begin{align}
f(r_h)&=0,&f'(r_h)=\frac{d-3}{r_h}, \label{eq.fBCs}
\end{align}
where $r_h$ is the outer horizon radius and (quasi)extremality corresponds to $r_h\rightarrow 0$. In terms of $z$ and $\chi$, the gauge condition is
\begin{align}
\chi(z_h)&=\frac 1{z_h},&\dot \chi(z_h)=0, \label{eq.chiBCs}
\end{align}
where $z_h=\frac1{(d-3) V_{d-2} r_h^{d-3}}$, and (quasi)extremality corresponds to $z_h\rightarrow \infty$.

Note that the above gauge choice can be restated as
\be
g_h A_h = \frac{1}{2z_h} \,, \label{eqn:gAzRelation}
\ee
relating the surface gravity $g_h$ and horizon area $A_h$ to the coordinate location of the horizon $z=z_h$.

\subsection{Self-force corrections \label{sec.eforce}}

We first observe that $T^r_r+\frac 1{d-2}T^i_i = \ap ( T_\text{hd}{}^r_r+\frac 1{d-2} T_\text{hd}{}^i_i)$ since the two-derivative action~\eqref{eq.twoderivaction} makes no contribution to this particular combination. Thus, using the boundary conditions~\eqref{eq.chiBCs} and the appropriate Green's function, we solve~\eqref{eq.chieqn} to obtain
\be
\chi(z)=\frac 1{z_h}-\frac{2 \ap k_N}{d-3} \int_{z_h}^z\left(\frac z{z'}-1\right) e^{2 \psi(z')} {A}^2(z') \left((d-2)\Thd{}^r_r(z')+ \Thd{}^i_i(z')\right)dz' . \label{eq.chisoln}
\ee
In particular, due to the explicit appearance of $\ap$ in the second term, this equation fixes the order-$\ap$ correction to $\chi(z)$ in terms of the functional form of $\Lhd$ along with the leading-order fields.

To relate this to the long-range part of the force between identical electrically-charged black holes, note that the latter takes the form~\cite{Heidenreich:2020upe}
\begin{align}
\mathbf{F}_\text{self}(r) &= \frac{\Fself}{V_{d-2}} \frac{\mathbf{\hat{r}}}{r^{d-2}}+ \ldots , &
\Fself &= f^{AB}_\infty Q_A Q_B-G^{ab}_\infty\mu_a\mu_b - k_N M^2,\label{eq.Fselfdef}
\end{align}
where $f^{AB}_\infty = f^{AB}(\phi_\infty)$, $G^{ab}_\infty = G^{ab}(\phi_\infty)$ for $\phi_\infty^a = \phi^a(r=\infty)$ the vacuum at spatial infinity, $M$ is the mass of the black hole, and $\mu_a$ is the ``scalar charge", 
\be
\mu_a\equiv \frac{\partial M}{\partial \phi_\infty^a}, \label{eq.muder}
\ee
i.e., the derivative of the mass of the black hole with respect to the values of the scalar field at spatial infinity. Equivalently, $\mu_a$ determines the long-range behavior of the scalar fields,
\begin{equation}
\phi^a =  \phi_\infty^a - \frac{G_\infty^{a b} \mu_b}{(d-3)V_{d-2} r^{d-3}}+ \ldots \label{eq.mu}
\end{equation}
so that $\mu_a = - G_{a b}^\infty \dot\phi^b_\infty$.\footnote{The derivation of this relation between the scalar charge and the long range scalar fields can be found in, e.g., \cite{Heidenreich:2020upe} \S4.1, at the two-derivative level. In fact, the argument is unchanged by derivative corrections, whose contributions to the probe particle action are velocity-dependent and whose contributions to the linearized backreaction fall off more rapidly than the leading-order $1/r^{d-3}$ contributions.\label{fn:scalarcharge}}

Evaluating~\eqref{eq.conseqn} at spatial infinity, we obtain
\be
\dot\psi_\infty\left[\dot \psi_\infty-\chi_\infty\right]-\dot \chi_\infty=k_N f_\infty^{A B} Q_A Q_B - k_N G^\infty_{a b} \dot{\phi}^a_\infty \dot{\phi}^b_\infty . \label{eq.consinf}
\ee
Note that the contributions to~\eqref{eq.conseqn} involving $\ap  T_\text{hd}{}_r^r$ have all dropped out because they invariably fall off too quickly as $r \to \infty$.

Using, e.g., the formulae in \cite{Lu:1993vt}, we obtain the ADM mass
\be
M=\frac{1}{k_N}\left(\frac12\chi_\infty-\dot \psi_\infty\right) . \label{eq.ADMmass}
\ee
Thus,~\eqref{eq.consinf} becomes
\be
\Fself =f_\infty^{A B} Q_A Q_B-G^{ab}_\infty\mu_a\mu_b-k_N M^2=-\frac{1}{k_N}\left(\dot \chi_\infty+\frac14\chi_\infty^2\right).
\ee
Specializing to the (quasi)extremal case, we obtain
\begin{subequations}
\begin{align}
\chi_\infty &=-\frac{2 \ap k_N}{d-3}\int_{0}^\infty e^{2\psi} A^2 \left((d-2)\Thd{}^r_r+ \Thd{}^i_i \right)dz , \label{eq.chiInfty} \\
\dot\chi_\infty &=\frac{2 \ap k_N}{d-3}\int_{0}^\infty \frac{e^{2\psi}A^2}{z} \left((d-2)\Thd{}^r_r+ \Thd{}^i_i \right)dz ,
\end{align}
\end{subequations}
using \eqref{eq.chisoln}. Thus, the self-force coefficient of a quasiextremal solution is
\be
\Fself=-2 \ap V^2_{d-2}  \int_0^\infty  \Bigl((d-2)\Thd{}^r_r+ \Thd{}^i_i \Bigr) e^{-\frac{2\psi}{d-3}} r^{2d-5} dr + O(\ap^2).\label{eq.Fselfsolpre}
\ee
This vanishes at leading order in the derivative expansion, as first shown in \cite{Heidenreich:2020upe}. Using $\mathcal R=re^{-\frac\psi{d-3}}$ from \eqref{eq.ansatzfromintro}, we can rewrite the force formula in the alternative way:
\begin{equation}
	\Fself=-2 \ap V^2_{d-2}  \int_{r_h}^\infty  \Bigl((d-2)\Thd{}^r_r+\Thd{}^i_i \Bigr) \mathcal{R}^{2d-5} |g_{t t}| \sqrt{g_{r r}} dr+O(\ap^2).\label{eq.Fselfsol}
\end{equation}

\subsection{Mass corrections \label{sec.emass}}

Let $\mathcal{M}(\phi)$ be the mass of an extremal black hole of fixed charge at two-derivative order, expressed as a function of the asymptotic values of the moduli, and define
\be
Q^2(\phi)\equiv f^{AB}(\phi)Q_AQ_B.
\ee
This mass function $\mathcal M(\phi)$ satisfies the condition
\be
Q^2(\phi)= k_N \mathcal M^2(\phi)+ G^{ab}(\phi)\mathcal M_{,a}(\phi)\mathcal M_{,b}(\phi), \label{eq.fakeW} 
\ee
related to the vanishing of the long-range self-force at two-derivative order. Moreover, the two-derivative extremal solution is an $\mathcal{M}(\phi)$ gradient flow, solving
\begin{align}
\dot\psi &= - k_N e^\psi \mathcal{M} , & \dot \phi^a &= - e^\psi G^{a b} \mathcal{M}_{,b} . \label{eq.gradflow}
\end{align}
The function $\mathcal{M}(\phi)$ is also known as the ``fake-superpotential''~\cite{Ceresole:2007wx,Andrianopoli:2007gt,Andrianopoli:2009je,Andrianopoli:2010bj,Trigiante:2012eb}; it can be calculated systematically by solving~\eqref{eq.fakeW}, see, e.g., \cite{Harlow:2022gzl} for a review.

To quantify the change in the solution due to derivative corrections, we define
\begin{align}
X& \equiv  f\dot\psi+ k_N \sqrt f e^\psi \mathcal M,&Y^a &\equiv  f \dot \phi^a+ \sqrt f e^\psi G^{ab}\mathcal M_{,b}, \label{eq.XYdefn}
\end{align}
where the particular powers of $f$ are chosen for future convenience. Since extremal solutions satisfy $f=1$ at two-derivative order, $X=Y^a=0$ for extremal two-derivative solutions per~\eqref{eq.gradflow}. Thus, for extremal derivative-corrected solutions, $X$ and $Y^a$ are $O(\ap)$. Eliminating $\dot{\psi}_\infty$ in favor of $X_\infty$, the ADM mass~\eqref{eq.ADMmass} becomes:
\be
M=\mathcal{M} + \frac{1}{k_N}\biggl(\frac12\chi_\infty- X_\infty\biggr) .
\ee
Thus, $\delta M = \frac{1}{k_N}\left(\frac12\chi_\infty- X_\infty\right)$ evaluated on a derivative-corrected extremal solution is the extremal mass correction we are interested in.

To determine this combination, consider the $tt$ component of the Einstein equations (a linear combination of \eqref{eq.psieqn} and \eqref{eq.conseqn}), which takes the form:
\be
\frac{1}{2} \dot \chi -\frac{d}{dz}\bigl(f\dot\psi\bigr) + \frac{1}{2} \dot\psi\Bigl[f\dot \psi+\dot f\Bigr] = k_N e^{2 \psi} A^2\, T^t_t .
\ee
Eliminating $\dot{\psi}$ in favor of $X$, we obtain
\begin{equation}
\frac{1}{2} \dot \chi - \dot{X}
 + \frac{X^2 + X \dot{f}}{2 f} = \frac{1}{2} k_N^2 e^{2\psi} \mathcal{M}^2 - k_N \sqrt f e^\psi \mathcal M_{,a}\dot{\phi}^a+ k_N e^{2 \psi} A^2\, T^t_t .
\end{equation}
Rewriting $\mathcal{M}^2$ using~\eqref{eq.fakeW} and then eliminating $\mathcal{M}_{,a}$ in terms of $Y^a$ and $\dot{\phi}^a$ using~\eqref{eq.XYdefn}, we find:
\begin{multline}
 \frac{1}{2} \dot \chi - \dot{X} +
 \frac{X^2 + X \dot{f}+k_N G_{ab} Y^a Y^b}{2 f}  = \frac{1}{2} k_N (e^{2\psi} Q^2 + f G_{ab} \dot{\phi}^a \dot{\phi}^b) + k_N e^{2 \psi} A^2\, T^t_t .
\end{multline}
To make use of this expression, note that the last term on the left-hand-side is $O(\ap^2)$ for an extremal solution. Since $X$ vanishes on the horizon,\footnote{This follows from $f(z_h) = 0$ in the nonextremal case and from $f \to \text{finite}$ and $z e^{\psi} \to \text{finite}$ as $z\to \infty$ in the quasiextremal case.} as does $\chi$ in the quasiextremal case, we conclude that
\be
\delta M = \frac{1}{k_N}\biggl[\frac12\chi_\infty- X_\infty\biggr] = -\int_{0}^\infty \biggl[\frac{1}{2} e^{2\psi} Q^2 + \frac{1}{2} f G_{ab} \dot{\phi}^a \dot{\phi}^b + e^{2 \psi} A^2\, T^t_t \biggr] d z + O(\ap^2) . \label{eq-mass-penultimate}
\ee
The first two terms on the right-hand-side cancel the leading order contributions to $T^t_t$, leaving:
\be
\delta M =  - \ap V_{d-2}  \int_{0}^\infty \biggl(\Thd{}^t_t + F^A_{tr} \frac{\delta \Shd}{\delta F^A_{tr}} \biggr) e^{-\frac{2\psi}{d-3}} r^{d-2} dr + O(\ap^2) . \label{eq.masspre}
\ee

Using $\mathcal R=re^{-\frac\psi{d-3}}$ from \eqref{eq.ansatzfromintro}, we can rewrite the mass formula in the alternative way that suggests a covariant generalization:
\begin{equation}
\delta M=- \ap V_{d-2}  \int_{r_h}^\infty \biggl(\Thd{}^t_t + F_{t r}^A \frac{\delta \Shd}{\delta F^A_{tr}} \biggr) \mathcal{R}^{d-2} \sqrt{|g_{t t} g_{r r}|} dr + O(\ap^2).\label{eq.mass}
\end{equation}
This matches a covariant formula that was derived for Reissner-Nordstr\"om black holes (i.e., without moduli) in \cite{Aalsma:2020duv}, compared with which our result is both more general (allowing for arbitrary moduli) and less general (requiring spherical symmetry).

\subsection{Entropy corrections \label{sec.eentropy}}
Corrections to black hole entropies induced by higher-derivative operators were previously studied in certain contexts in, e.g., \cite{Sen:2005iz,Sahoo:2006pm,Cheung:2018cwt,Cremonini:2019wdk}. In this subsection, we use the attractor mechanism  \cite{Ferrara:1995ih,Cvetic:1995bj,Strominger:1996kf,Ferrara:1996dd,Ferrara:1996um} to compute the entropy correction to spherically-symmetric extremal black holes in general effective field theories with moduli.

The Iyer-Wald entropy $\mathcal{S}$~\cite{Iyer:1994ys} is defined as 
\be
\mathcal{S}\equiv -2\pi\int_\Sigma \frac{1}{4} \frac{\delta S}{\delta R_{\mu \nu \rho \sigma}}\epsilon_{\mu \nu} \epsilon_{\rho \sigma} dA_{d-2}, \label{eq.entropydef}
\ee
where $\Sigma_h$ is the event horizon, $\epsilon_{\mu \nu}$ is binormal to $\Sigma_h$ with $\epsilon_{\mu \nu}\epsilon^{\mu \nu}=-2$, $dA_{d-2}$ is the volume-form on the event horizon, and the functional derivative with respect to the Riemann tensor is defined by
\be
\delta S = \int d^d x \sqrt{-g} \frac{1}{4} \frac{\delta S}{\delta R_{\mu \nu \rho \sigma}} \delta R_{\mu \nu \rho \sigma} \,,
\ee
where $\frac{\delta S}{\delta R_{\mu \nu \rho \sigma}}$ has the same symmetries as the Riemann tensor.

In our gauge 
$\epsilon_{\mu \nu}=\sqrt{-g_{tt}g_{rr}}\left(\delta^t_\mu \delta^r_{\nu}-\delta^r_\mu \delta^t_\nu\right)$.
Thus, performing the integral using spherical symmetry we obtain
\be
\mathcal{S} = 2 \pi A_h \frac{\delta S}{\delta R^{t r}\!_{t r}} \biggr|_{r=r_h} \,,
\ee
where $A_h$ is the area of the event horizon. In particular, at two-derivative order,
\begin{equation}
	\frac{\delta S_2}{\delta R^{m n}\!_{p q}}=\frac{1}{\kappa^2} (\delta^p_m \delta^q_n - \delta^p_n \delta^q_m),
\end{equation}
due to the Einstein-Hilbert action $S_2=\frac1{2\kappa^2} \int d^d x \sqrt{-g} R+\ldots$, so at leading order,
\begin{equation}
	\mathcal{S}^{(0)}=\frac{2\pi A_h^{(0)}}{\kappa^2},
\end{equation}
where $A_h^{(0)}$ is the leading-order horizon area.

Continuing to the next order, one finds
\be
\mathcal{S}=\mathcal{S}^{(0)}+ \frac{2\pi}{\kappa^2} \delta A_h+2\pi \ap A^{(0)}_h \frac{\delta \Shd}{\delta R^{tr}\!_{tr}}\biggr|_{r=r_h} + O(\alpha'^2), \label{eq.SSplit}
\ee
so there are two types corrections: (1) those arising from derivative corrections to the horizon area $\delta A_h$ and (2) those arising from operators in the higher-derivative action that involve the Riemann tensor.

\subsubsection*{The area correction via the attractor mechanism} 

To find the area correction $\delta A_h$ we use the attractor mechanism~\cite{Ferrara:1995ih,Cvetic:1995bj,Strominger:1996kf,Ferrara:1996dd,Ferrara:1996um}. Define
\be
x(z)\equiv \psi(z)+\log z.
\ee
The area of the horizon is related to $x_h\equiv \lim_{z\rightarrow \infty}x(z)$ by (see also~\eqref{eq.Aofz})
\begin{equation}
	A_h=V_{d-2}[(d-3)V_{d-2}e^{x_h}]^{-\frac{d-2}{d-3}}.\label{eq.Ahofxh}
\end{equation}
In terms of $x$, the uncorrected versions of \eqref{eq.phieqn} and \eqref{eq.psieqn} are
\begin{subequations} \label{eq.2dereom}
\begin{align}
\ddot \phi^a+\Gamma^a_{bc}\dot \phi^b\dot \phi^c&=\frac 1{2z^2}G^{ab}(\phi) Q^2_{,b}e^{2x},&&\text{(when no corrections),} \label{eq.2derddotphi}
\\
\ddot x&=\frac 1{z^2}\left(k_N Q^2e^{2x}-1\right),&&\text{(when no corrections).} \label{eq.2derddotx}
\end{align}
\end{subequations}
Looking back at the infalling metric~\eqref{eq.infall}, we see that a smooth, extremal horizon requires $x$ and $f$ to be smooth functions of $\rho = r^{d-3}$ at $\rho = 0$, and $\phi^a$ must be as well if the moduli are smooth at the horizon. Expressed in terms of $z \propto 1/\rho$, any such function $F(z)$ must have a finite limit $F_h \equiv \lim_{z\to \infty} F(z)$, whereas its $n$th derivative $F^{(n)}(z)$ must fall off faster than $1/z^n$.

In particular $x_h\equiv \lim_{z\rightarrow \infty}x(z)$ must be finite, whereas $z \dot x$ and $z^2 \ddot x$ tend to zero as $z \to \infty$, and likewise for $\phi^a_h$, $z \dot \phi^a$ and $z^2 \ddot \phi^a$. Thus, multiplying \eqref{eq.2dereom} by $z^2$ and taking the $z\rightarrow \infty$ limit,
\begin{align}
Q^2_{,a}(\phi_h) &=0,& 
 Q^2(\phi_h) &=(k_N e^{2x_h})^{-1},&&\text{(when no corrections).} \label{eq.2derAttract}
\end{align}
These equations both fix the values of the moduli at the event horizon $\phi_h^a$ (with some ambiguity when $Q^2(\phi)$ has multiple critical points) and also determine the horizon area of the resulting extremal solution.

We now examine how derivative corrections modify this attractor argument. In terms of $x$, the equations of motion \eqref{eq.zeom} can be rewritten as
\begin{subequations}\label{eq.xeom}
	\begin{align}
z^2 \frac{d}{d z}  (f \dot{\phi}^a) + f \Gamma^a_{b c} z \dot{\phi}^b z \dot{\phi}^c &= e^{2 x} A^2 G^{a b} \biggl( \frac{1}{2} f_{A B, b} F^A \cdot F^B - \alpha'  \frac{\delta \Shd}{\delta \phi^b} \biggr),\\
z^3 \frac{d^2}{d z^2}  \biggl( \frac{1 - f}{z} \biggr) &= - \frac{2k_N e^{2 x}  A^2}{d-3}   \bigl( (d-2)T_r^r +  T^i_i \bigr), \\
z^2 \frac{d}{d z}  \biggl( f \dot{x} - \frac{1}{z} \biggr) &= -k_N e^{2 x} A^2  (T^t_t + T_r^r), \\
f \bigl(z^2 \ddot{x} - z^2 \dot{x}^2+2 z \dot{x}\bigr) &= k_N e^{2 x} A^2  (T_r^r - T^t_t) ,
\end{align}
\end{subequations}
where the last equation is a linear combination of \eqref{eq.psieqn} and \eqref{eq.conseqn}. Consider the $z \rightarrow \infty$ limit of the equations \eqref{eq.xeom}, requiring that $f \rightarrow f_h$, $x \rightarrow x_h$, and $\phi^a \rightarrow \phi^a_h$, with $n$th derivatives of these quantities falling off faster than $1 / z^n$. After a bit of rearranging, we obtain the following attractor equations
\begin{subequations}\label{eq.attractor}
\begin{align}
Q^2_{,a} (\phi_h) &= 2 \alpha' A_h^2 \biggl[ \frac{\delta \Shd}{\delta \phi^a} + f_{A B} f^{A C}_{, a} F^B_{t r} \frac{\delta \Shd}{\delta F^C_{t r}} \biggr]_{r = r_h}, \label{eq.attractorQder} \\
f_h &= 1 + \alpha' \kappa^2 e^{2 x_h} A^2_h  \biggl[  T_\text{hd}{}_r^r + \frac{1}{d - 2}  T_\text{hd}{}_i^i \biggr]_{r=r_h}, \label{eq.attractorfh}\\
Q^2 (\phi_h) &= \frac{1}{k_N e^{2 x_h}} + \alpha' A^2_h  \left[ T_\text{hd}{}_t^t + T_\text{hd}{}_r^r + 2 F^A_{t r}\frac{\delta \Shd}{\delta F^A_{t r}} \right]_{r=r_h},\label{eq.attractorQ} \\
0 &= \bigl[ T_\text{hd}{}_{r}^r -  T_\text{hd}{}^t_{ t} \bigr]_{r=r_h}. \label{eq.attractorTrrTtt}
\end{align}
\end{subequations}
Equation \eqref{eq.attractorQ} tells us how $x_h$, which is related to the area of the black hole by \eqref{eq.Ahofxh}, depends on the value of $Q^2(\phi)$ at the horizon. Meanwhile, \eqref{eq.attractorQder} governs the values that the moduli must take at the event horizon. Note that while \eqref{eq.attractorfh} likewise fixes $f_h$ (which is not needed in our present calculation), \eqref{eq.attractorTrrTtt} at first glance appears to constrain the two-derivative solution itself in a manner that depends on the higher-derivative corrections. In fact, \eqref{eq.attractorTrrTtt} is identically true because the near-horizon geometry of a large black hole at two-derivative order is AdS$_2 \times S^{d-2}$, and the symmetries thereof require $T^\mu_\nu \propto \delta^\mu_\nu$ along the AdS$_2$.

In principle, the corrected horizon area is determined by first solving~\eqref{eq.attractorQder} to determine the values of the moduli at the horizon $\phi_h^a = (\phi_h^a)^{(0)} + \delta\phi_h^a$, then substituting these values into~\eqref{eq.attractorQ} to fix $x_h$, then applying~\eqref{eq.Ahofxh} to obtain the horizon area $A_h$. However, Taylor expanding $Q^2(\phi_h)$ about the leading-order attractor point $\phi_h^{(0)}$, one finds
\be
Q^2(\phi_h) 
 = Q^2(\phi_h^{(0)}) + \frac{1}{2} \delta \phi^a \delta \phi^b Q^2_{,a b}(\phi_h^{(0)}) + \ldots \,,
\ee
where terms linear in $\delta\phi^a$ vanish due to the leading-order attractor equation $Q^2_{,a}(\phi_h^{(0)}) = 0$, see~\eqref{eq.2derAttract}. Thus, since $\delta\phi^a$ is $O(\alpha')$ per~\eqref{eq.attractorQder}, the leading correction to $Q^2(\phi_h)$ is $O(\alpha'^2)$. As a consequence, expanding~\eqref{eq.attractorQ} to linear order in $\ap$ yields
\be
0 = -\frac{2}{k_N e^{2 x_h}} \delta x_h + \alpha' A^2_h  \biggl[ \Thd{}_t^t + \Thd{}_r^r + 2 F^A_{t r}\frac{\delta \Shd}{\delta F^A_{t r}} \biggr]_{r=r_h} + O(\ap^2) \,.
\ee
Applying~\eqref{eq.Ahofxh} to eliminate $x_h$ in favor of $A_h$, we obtain
\be
\frac{\delta A_h}{A_h} =-\frac{d-2}{d-3} \delta x_h = -\frac{\ap \kappa^2 \mathcal{R}_h^2}{(d-3)^2} \biggl[ \Thd{}_t^t + F^A_{t r}\frac{\delta \Shd}{\delta F^A_{t r}} \biggr]_{r=r_h} + O(\ap^2) ,
\ee
where $\mathcal{R}_h = \bigl[\frac{A_h}{V_{d-2}}\bigr]^{\frac{1}{d-2}}$ is the curvature radius of the horizon and we use~\eqref{eq.attractorTrrTtt} to eliminate $\Thd{}_r^r$ in favor of $\Thd{}_t^t$.

Substituting this into~\eqref{eq.SSplit}, we obtain the extremal entropy correction
\be
\delta\mathcal{S}= 2 \pi \ap V_{d-2} \biggl[ -\frac{\mathcal{R}^d}{(d-3)^2} \biggl(\Thd{}_t^t + F^A_{t r}\frac{\delta \Shd}{\delta F^A_{t r}}\biggr) + \mathcal{R}^{d-2} \frac{\delta \Shd}{\delta R^{tr}\!_{tr}}\biggr]_{r = r_h}+ O(\alpha'^2) .\label{eq.preentropy}
\ee

\subsection{The dyonic case \label{sec.dcorrections}}

In four dimensions, static, spherically-symmetric black holes can be dyonic, carrying both electric and magnetic charge. In our preceding analysis, we assumed that only electric charge was present. We now examine the four-dimensional dyonic case, showing that our final results \eqref{eq.Fselfsol}, \eqref{eq.mass}, and \eqref{eq.preentropy} are unchanged.

Firstly, because of the presence of magnetic charge, we can no longer neglect moduli-dependent $\theta$ terms in the two-derivative effective action of the form
\be
S_\theta = \frac{1}{8\pi^2} \int \theta_{A B}(\phi) F^A \wedge F^B \,.
\ee
Accounting for such $\theta$ terms, the gauge-field equations of motion become
\begin{align}
d\star \mathcal F_A&=0,&d F^A&=0, &
\text{where}\quad
\mathcal F_A&=f_{AB}F^B + \theta_{AB} \star F^B - \ap \frac{\delta \Shd}{\delta F^A}. \label{eq.gaugeeom-dyonic}
\end{align}
The conserved electric and magnetic charges are thus\footnote{To be precise, these are the Page charges (see, e.g.,~\cite{Marolf:2000cb}), which are quantized and conserved, but not invariant under large gauge transformations (in this case, constant shifts of $\theta_{A B}(\phi)$ by amounts that leave the quantum theory unchanged).}
\begin{align}
Q_A^{(e)} &= \oint_{S^{2}} \star \mathcal F_A , & Q^A_{(m)} &= \frac{1}{2\pi}\oint_{S^{2}} F^A .
\end{align}
Spherical symmetry then implies that
\begin{subequations}
\begin{align}
\star \mathcal F_A &= (\star \mathcal F_A)_{t r} d t \wedge d r + Q_A^{(e)} \frac{\sin \theta d\theta \wedge d \varphi}{4 \pi} , \\
F^A &= F^A_{t r} d t \wedge d r + Q^A_{(m)} \frac{\sin \theta d\theta \wedge d \varphi}{2} ,
\end{align}
\end{subequations}
where $\theta$, $\varphi$ are the standard coordinates on $S^2$ and $V_2 = 4\pi$ is its volume.
Eliminating $\mathcal{F}_A$ in favor of $F^A$, we obtain
\be
F^A = f^{A B} \biggl[ -\biggl(Q_B^{(e)} + \frac{\theta_{B C}}{2\pi} Q^C_{(m)}\biggr) \frac{e^{2\psi}}{4 \pi r^2} + \ap \frac{\delta \Shd}{\delta F^{B t r}} \biggr] dt \wedge d r + Q_{(m)}^A \frac{\sin \theta d \theta \wedge d \varphi}{2} \,.
\ee
Note that this reduces to~\eqref{eq.F} upon setting $Q^A_{(m)} = 0$.

Apart from this modification to the form of $F^A_{\mu \nu}$, the Einstein equations are unchanged from before---since the $\theta$ terms do not couple to the metric---whereas the moduli equations of motion become
\be
\frac d{dz}(f\dot\phi^a)+f\Gamma^a_{bc}\dot\phi^b\dot\phi^c = e^{2 \psi} A^2\, G^{ab} \biggl(\frac{1}{2} f_{AB,b}F^A\cdot F^B + \frac{1}{8\pi^2} \theta_{AB,b}F^A\cdot \star F^B -\ap \frac{\delta \Shd}{\delta \phi^b} \biggr), 
\ee
rather than~\eqref{eq.phieqn}.

We can then proceed exactly as before until we reach~\eqref{eq.consinf}, which now reads
\be
\dot\psi_\infty\left[\dot \psi_\infty-\chi_\infty\right]-\dot \chi_\infty=k_N Q^2(\phi) - k_N G^\infty_{a b} \dot{\phi}^a_\infty \dot{\phi}^b_\infty ,
\ee
where
\be
Q^2(\phi) \equiv f^{A B}(\phi) \biggl[Q_A^{(e)} + \frac{\theta_{A C}(\phi)}{2\pi} Q_{(m)}^C \biggr] \biggl[Q_B^{(e)} + \frac{\theta_{B D}(\phi)}{2\pi} Q_{(m)}^D \biggr] + 4\pi^2f_{A B}(\phi) Q_{(m)}^A Q_{(m)}^B \,.
\ee
The extra terms are precisely those appearing in the coefficient of the self-force
\be
\Fself = Q^2(\phi_\infty)-G^{ab}_\infty\mu_a\mu_b - k_N M^2 \,,
\ee
see, e.g.,~\cite{Heidenreich:2020upe} for details, so we still obtain
\begin{equation}
	\Fself=-2 \ap V^2_{2}  \int_{r_h}^\infty  \Bigl(2\Thd{}^r_r+\Thd{}^i_i \Bigr) \mathcal{R}^{3} |g_{t t}| \sqrt{g_{r r}} dr+O(\ap^2) ,
\end{equation}
in an identical manner to before.

Likewise, the calculation of the mass correction proceeds identically through~\eqref{eq-mass-penultimate}. To obtain our final answer from here, we plug in the explicit form of $Q^2(\phi)$ and $T^t_t$; each has extra terms in the dyonic case, but these extra terms fortuitously cancel,\footnote{This cancellation can be traced back to the fact that the two-derivative portions of $T^t_t$ and $T^r_r$ depend on $F^A$ in exactly the same way.} leading once again to
\begin{equation}
\delta M=- \ap V_{2}  \int_{r_h}^\infty \biggl(\Thd{}^t_t + F_{t r}^A \frac{\delta \Shd}{\delta F^A_{tr}} \biggr) \mathcal{R}^{2} \sqrt{|g_{t t} g_{r r}|} dr + O(\ap^2).
\end{equation}
Note that, unlike in the electric case, $F_{t r}^A \frac{\delta \Shd}{\delta F^A_{tr}} \ne \frac{1}{2} F_{\mu \nu}^A \frac{\delta \Shd}{\delta F^A_{\mu \nu}}$, so it is important to write the formula in this particular form.

The entropy calculation is likewise virtually unchanged, and we again find
\be
\delta\mathcal{S}=  - 2 \pi \ap V_{2} \mathcal{R}_h^4 \biggl[ \Thd{}_t^t + F^A_{t r}\frac{\delta \Shd}{\delta F^A_{t r}} + R^{t r}\!_{tr} \frac{\delta \Shd}{\delta R^{tr}\!_{tr}}\biggr]_{r = r_h}+ O(\alpha'^2), 
\ee
just as before.

\subsection{Further simplifications} \label{subsec:simplifications}

In fact, the mass and entropy correction formulas \eqref{eq.mass} and \eqref{eq.preentropy} can be further simplified, as follows. First, notice that for
the independent three and four-derivative operators classified in appendix~\ref{sec.basis} (see also appendix~\ref{app:stresstensor} for useful formulas),
$\Thd{}^t_t + F_{t r}^A  \frac{\delta \Shd}{\delta F_{t r}^A}$
is always equal to the Lagrangian density $\Lhd$ when
evaluated in a spherically symmetric background, \textbf{except} for
operators involving the Riemann tensor.

To generalize this observation, we begin by assuming that
$\Lhd$ depends on the metric, $\nabla_{\mu} \phi^a$,
$F_{\mu \nu}^A$, and $R^{\mu}_{\; \nu \rho \sigma}$, but
\emph{not} on higher covariant derivatives thereof. 
 Thus, the variation in
$\Lhd$ as these constituents are varied is\footnote{Note that to define the partial derivative $\frac{\partial\Lhd}{\partial g_{\mu \nu}}$ we need not only specify that $\nabla_{\mu} \phi^a, F_{\mu \nu}^A$ and
$R^{\mu}_{\; \nu \rho \sigma}$ are held fixed, but also that~\eqref{eqn:Lvariation} holds with $\frac{\partial \Lhd}{\partial R_{\mu \nu \rho \sigma}} \equiv g^{\mu \alpha} \frac{\partial\Lhd}{\partial R^{\alpha}_{\; \nu \rho \sigma}}$ chosen to have all the symmetries of the Riemann tensor. The reason for this subtlety is that $g_{\mu \alpha}
\delta R^{\alpha}_{\; \nu \rho \sigma}$ \emph{does not} have all the symmetries of the Riemann tensor; in particular, $g_{(\mu| \alpha}
\delta R^{\alpha}_{\; |\nu) \rho \sigma} = -\delta g_{(\mu| \alpha} R^\alpha_{\; |\nu)\rho \sigma}$ since $\delta (R_{\mu \nu \rho \sigma}) = \delta (g_{\mu \alpha} R^{\alpha}_{\; |\nu) \rho \sigma})$ \emph{does} retain all the symmetries. Thus, adding a term symmetric in the exchange of $\mu,\nu$ to $\frac{\partial \Lhd}{\partial R_{\mu \nu \rho \sigma}}$ changes $\frac{\partial\Lhd}{\partial g_{\mu \nu}}$ without altering the dependence of $\Lhd$ on the Riemann tensor.} 
\begin{equation}
  \delta \Lhd = \frac{\partial
  \Lhd}{\partial g_{\mu \nu}} \delta g_{\mu \nu} +
  \frac{\partial \Lhd}{\partial \nabla_{\mu} \phi^a} \delta
  \nabla_{\mu} \phi^a + \frac{1}{2}  \frac{\partial
  \Lhd}{\partial F_{\mu \nu}^A} \delta F_{\mu \nu}^A +
  \frac{1}{4}  \frac{\partial \Lhd}{\partial
  R^{\mu}_{\; \alpha \beta \gamma}} \delta R^{\mu}_{\;
  \alpha \beta \gamma}, \label{eqn:Lvariation}
\end{equation}
where partial derivatives with respect to tensor fields are defined to enjoy
the same symmetries as the tensor field in question, and the factors of $1 /
2$ and $1 / 4$ reflect our normalization conventions (so that, e.g.,
$\frac{\partial F_{\mu \nu}^A}{\partial F_{\rho \sigma}^B} = \delta^A_B
(\delta_{\mu}^{\rho} \delta_{\nu}^{\sigma} - \delta_{\nu}^{\rho}
\delta_{\mu}^{\sigma})$). In particular, due to general covariance\footnote{Here we implicitly exclude gravitational Chern-Simons terms, as justified in appendix~\ref{sec.basis}.}
 $\Lhd$ must be
invariant under an infinitesimal coordinate transformation $x^\mu \to x^\mu - \varepsilon^\mu_\nu x^\nu + \cdots$ for any $(1, 1)$ tensor field $\varepsilon^{\mu}_{\nu}$, resulting in
\begin{align}
  \delta g_{\mu \nu} & = \varepsilon^{\rho}_{\mu} g_{\rho \nu} +
  \varepsilon^{\rho}_{\nu} g_{\mu \rho}, \quad \delta \nabla_{\mu} \phi^a =
  \varepsilon_{\mu}^{\nu} \nabla_{\nu} \phi^a , \nonumber\\
  \delta F_{\mu \nu}^A & = \varepsilon_{\mu}^{\rho} F_{\rho \nu}^A +
  \varepsilon_{\nu}^{\rho} F_{\mu \rho}^A, \quad \delta
  R^{\mu}_{\; \alpha \beta \gamma} = - \varepsilon^{\mu}_{\rho}
  R^{\rho}_{\; \alpha \beta \gamma} + \varepsilon_{\alpha}^{\rho}
  R^{\mu}_{\; \rho \beta \gamma} + \varepsilon_{\beta}^{\rho}
  R^{\mu}_{\; \alpha \rho \gamma} + \varepsilon_{\gamma}^{\rho}
  R^{\mu}_{\; \alpha \beta \rho}, 
\end{align}
which applied to~\eqref{eqn:Lvariation} implies the identity
\begin{equation}
  0 = 2 g_{\nu \rho} \frac{\partial \Lhd}{\partial g_{\mu
  \rho}} + \nabla_{\nu} \phi^a  \frac{\partial
  \Lhd}{\partial \nabla_{\mu} \phi^a} + F_{\nu \rho}^A 
  \frac{\partial \Lhd}{\partial F_{\mu \rho}^A} +
  \frac{1}{2} R^{\mu}_{\; \alpha \beta \gamma}  \frac{\partial
  \Lhd}{\partial R^{\nu}_{\; \alpha \beta
  \gamma}} . \label{eqn:CovarIdent}
\end{equation}

Next, we express the functional derivatives of $\Shd$ in terms of partial derivatives of $\Lhd$, as follows:
\begin{subequations}
\begin{gather}
  \frac{\delta \Shd}{\delta \nabla_{\mu} \phi^a}  = \frac{\partial
  \Lhd}{\partial \nabla_{\mu} \phi^a}, \qquad \frac{\delta
  \Shd}{\delta F_{\mu \nu}^A} = \frac{\partial
  \Lhd}{\partial F_{\mu \nu}^A}, \qquad \frac{\delta
  \Shd}{\delta R^{\mu}_{\; \alpha \beta \gamma}} =
  \frac{\partial \Lhd}{\partial R^{\mu}_{\;
  \alpha \beta \gamma}},  \label{eqn:partialToFunctional}\\
  \frac{\delta \Shd}{\delta g_{\mu \nu}}  = \frac{\partial
  \Lhd}{\partial g_{\mu \nu}} + \frac{1}{2} \nabla_{(\rho
  } \nabla_{ \sigma)} \frac{\partial
  \Lhd}{\partial R_{\rho \mu \nu \sigma}} + \frac{1}{2}
  g^{\mu \nu} \Lhd,  \label{eqn:metricFD}
\end{gather}
\end{subequations}
where to derive \eqref{eqn:metricFD}, take $\delta \Shd =
\int d^d x \sqrt{- g}  \bigl[ \frac{\partial \Lhd}{\partial
g_{\mu \nu}} \delta g_{\mu \nu} + \frac{1}{4}  \frac{\partial
\Lhd}{\partial R^{\mu}_{\; \nu \rho \sigma}}
\delta R^{\mu}_{\; \nu \rho \sigma} + \frac{1}{2} g^{\mu \nu}
\delta g_{\mu \nu} \Lhd \bigr]$ and integrate the middle
term by parts twice using $\delta R^{\mu}_{\; \nu \rho \sigma} =
\nabla_{\rho} \delta \Gamma_{\; \sigma \nu}^{\mu} -
\nabla_{\sigma} \delta \Gamma_{\; \rho \nu}^{\mu}$ and $\delta
\Gamma^{\mu}_{\; \nu \rho} = \frac{1}{2} g^{\mu \lambda} 
(\nabla_{\nu} \delta g_{\rho \lambda} + \nabla_{\rho} \delta g_{\nu \lambda} -
\nabla_{\lambda} \delta g_{\nu \rho})$.\footnote{Explicitly,
$\begin{aligned}[t]
  \delta S_{\text{mid}} &= \frac{1}{2} \int d^d x \sqrt{- g}  \frac{\partial
  \Lhd}{\partial R^{\mu}_{\; \nu \rho \sigma}}
  \nabla_{\rho} \delta \Gamma^{\mu}_{\; \sigma \nu} = - \frac{1}{2} \int
  d^d x \sqrt{- g} \nabla_{\rho} \frac{\partial
  \Lhd}{\partial R^{\mu}_{\; \nu \rho \sigma}}
  \delta \Gamma^{\mu}_{\; \sigma \nu} \nonumber\\
  &= - \frac{1}{2} \int d^d x \sqrt{- g} \nabla_{\rho} \frac{\partial
  \Lhd}{\partial R_{\mu \nu \rho \sigma}} \nabla_{\nu}
  \delta g_{\sigma \mu} = \frac{1}{2} \int d^d x \sqrt{- g} \nabla_{\nu}
  \nabla_{\rho} \frac{\partial \Lhd}{\partial R_{\mu \nu
  \rho \sigma}} \delta g_{\sigma \mu}.
\end{aligned}$}

Combining \eqref{eqn:CovarIdent}, \eqref{eqn:partialToFunctional},
(\ref{eqn:metricFD}) and the definition of the stress tensor $\Thd^{\mu
\nu} = 2 \frac{\delta \Shd}{\delta g_{\mu \nu}}$, we
obtain:
\begin{equation}
  \Thd{}^{\mu}_{\nu} + \nabla_{\nu} \phi^a  \frac{\delta
  \Shd}{\delta \nabla_{\mu} \phi^a} + F_{\nu \rho}^A  \frac{\delta
  \Shd}{\delta F_{\mu \rho}^A} + \frac{1}{2} R^{\mu}_{\;
  \alpha \beta \gamma}  \frac{\delta \Shd}{\delta
  R^{\nu}_{\; \alpha \beta \gamma}} + \nabla_{(\rho }
  \nabla_{ \sigma)} \frac{\delta \Shd}{\delta
  R^{\nu}_{\; \rho \mu \sigma}} = \delta^{\mu}_{\nu}
  \Lhd . \label{eqn:TLreln}
\end{equation}
In the special case of spherical symmetry and the absence of Riemann
couplings, we see that this reproduces $\Thd{}^t_t +
F_{t r}^A  \frac{\delta \Shd}{\delta F_{t r}^A}
= \Lhd$ as previously noted.

We now apply the relation \eqref{eqn:TLreln} to simplify the entropy
correction formula \eqref{eq.preentropy}. Observe that in the near horizon limit,
$\phi^a$, $F_{\mu \nu}^A$ and $R^{\mu}_{\; \nu \rho \sigma}$ are
all \emph{covariantly constant}. Thus, \emph{all covariant
derivatives} of these quantities vanish, and we obtain
\begin{equation}
  \biggl[ \Thd{}^t_t + F_{t r}^A  \frac{\delta
  \Shd}{\delta F_{t r}^A} + R_{t r t r}  \frac{\delta
  \Shd}{\delta R_{t r t r}} \biggr]_{r = r_h} = 
  \Lhd |_{r = r_h}.
\end{equation}
In fact, even though we derived this formula by assuming the absence of higher
covariant derivatives in $\Lhd$, it easily generalizes to
include such terms because all the covariant derivatives evaluate to zero in
the near-horizon limit, as already noted.

Thus, using $R^{t r}\!_{tr} = -\frac{(d-3)^2}{\mathcal{R}^2}$ in the near-horizon limit (see appendix~\ref{app:Riemann}), we obtain the general result
\begin{equation}
  \delta \mathcal{S}= \left. - \frac{2 \pi \alpha'}{(d - 3)^2} V_{d - 2}
  \mathcal{R}^d \Lhd \right|_{r = r_h} .
  \label{eqn:SLhdformula}
\end{equation}
The simplicity of this answer---in contrast with the complexity of its
derivation---suggests that there is a more general principle at work. However,
we leave further consideration of this to future work.

Next, we consider the mass formula \eqref{eq.mass}. Defining the projection tensor
$\Pi^{\mu}_{\nu} = \delta^{\mu}_t \delta^t_{\nu}$, we can write the integral
more coviariantly as
\begin{equation}
  \delta M = - \alpha'  \int_{\Sigma} \Pi_{\mu}^{\nu}  \biggl( \Thd{}_{\nu}^{\mu} + F_{\nu \rho}^A  \frac{\delta \Shd}{\delta F_{\mu \rho}^A} \biggr) N \sqrt{h}\, d^{d - 1} x,
\end{equation}
where the integral is taken over a spatial slice $\Sigma$ from the horizon to
infinity, $h$ is the determinant of the spatial metric and $N = \sqrt{- 1 / g^{t t}}$ is the
lapse function. Applying \eqref{eqn:TLreln}, this becomes:
\begin{equation}
  \delta M = - \alpha'  \int_{\Sigma} \biggl[ \Lhd -
  \frac{1}{2} \Pi_{\mu}^{\nu} R^{\mu}_{\; \alpha \beta \gamma} 
  \frac{\delta \Shd}{\delta R^{\nu}_{\; \alpha \beta
  \gamma}} - \Pi_{\mu}^{\nu} \nabla_{\rho} \nabla_{\sigma}  \frac{\delta
  \Shd}{\delta S^{\nu}_{\; \rho \mu \sigma}} \biggr] N
  \sqrt{h}\, d^{d - 1} x,
\end{equation}
since $\Pi_{\mu}^{\nu} \nabla_{\nu} \phi^a = 0$ and $\Pi_{\mu \nu} = \Pi_{\nu
\mu}$ in a static background. Computing the second covariant derivatives of
$\Pi^{\mu}_{\nu}$ in an extremal black hole background, one finds that\footnote{As a shortcut, first verify that 
$\nabla_{[\mu } \Pi_{ \nu]}^{\rho} = \nabla_{[\mu } \psi \Pi^{\rho}_{ \nu]}$ by explicit computation. It immediately follows that
$\nabla^{[\rho } \nabla_{[\mu } \Pi^{\sigma]}_{ \nu]} =
 \nabla^{[\rho } \nabla_{[\mu} \psi \Pi^{ \sigma]}_{ \nu]} + \nabla_{[\mu} \psi \nabla^{[\rho } \psi \Pi^{\sigma]}_{ \nu]}$.
Then, using $\nabla_{\mu} \nabla_{\nu} \psi =
\partial_{\mu} \partial_{\nu} \psi - \Gamma_{\; \mu \nu}^{\rho}
\partial_{\rho} \psi$ and the explicit form of the connection (see appendix~\ref{app:Riemann}), one obtains $\nabla^{[t } \nabla_{[t }
\Pi^{ r]}_{ r]} = - \frac{1}{4} R^{t r}_{t r}$ and
$\nabla^{[t } \nabla_{[t } \Pi^{ i]}_{
j]} = - \frac{1}{4} R^{t i}_{t j}$, or equivalently $\nabla^{[\rho }
\nabla_{[\mu } \Pi^{ \nu]}_{ \sigma]} = -
\frac{1}{2} R^{\alpha [\nu }_{\mu \sigma} \Pi_{\alpha}^{
\rho]}$.} 
\begin{equation}
  X^{\; \rho \mu \sigma}_{\nu} \nabla_{\rho} \nabla_{\sigma}
  \Pi^{\nu}_{\mu} = - \frac{1}{2} \Pi_{\mu}^{\nu} R^{\mu}_{\;
  \alpha \beta \gamma} X^{\; \alpha \beta \gamma}_{\nu},
\end{equation}
for any $X_{\mu \nu \rho \sigma}$ with the symmetries of the Riemann tensor.
Thus, we obtain:
\begin{align}
  \delta M &= - \alpha'  \int_{\Sigma} \biggl[ \Lhd +
  \nabla_{\rho} \biggl( \nabla_{\sigma} \Pi^{\nu}_{\mu}  \frac{\delta
  \Shd}{\delta R^{\nu}_{\; \rho \mu \sigma}} \biggr) -
  \nabla_{\sigma} \biggl( \Pi^{\nu}_{\mu} \nabla_{\rho} \frac{\delta
  \Shd}{\delta R^{\nu}_{\; \rho \mu \sigma}} \biggr)
  \biggr] N \sqrt{h} d^{d - 1} x \nonumber\\
  &= - \alpha' \int_{\Sigma} \Lhd N \sqrt{h} d^{d - 1} x
  \left. - \alpha'  \hat{r}_{\alpha} \biggl[ \nabla_{\sigma} \Pi^{\nu}_{\mu} 
  \frac{\delta \Shd}{\delta R^{\nu}_{\; \alpha \mu
  \sigma}} - \Pi^{\nu}_{\mu} \nabla_{\rho} \frac{\delta \Shd}{\delta
  R^{\nu}_{\; \rho \mu \alpha}} \biggr] N\mathcal{R}^{d - 2}
  \right|_{r = r_h}^{\infty}, 
\end{align}
after converting the total derivatives into boundary terms,\footnote{Note that
the explicit presence of the lapse function in the measure compensates for the
fact that $\nabla_{\mu}$ is the \emph{spacetime} covariant derivative.}
where $\hat{r}_{\mu} = \sqrt{g_{r r}} \delta_{\mu}^r$ is the radially-outwards
unit vector. In fact, since the lapse function $N$ vanishes at the horizon and
the fields fall off sufficiently rapidly at infinity, the boundary terms
vanish, and we finally obtain\footnote{This formula appeared previously in~\cite{Cheung:2019cwi,Arkani-Hamed:2021ajd} in the four-dimensional case without moduli (but allowing for rotation).}
\begin{equation}
  \delta M = - \alpha' \int_{\Sigma} \Lhd N \sqrt{h}\, d^{d-1} x = - \alpha' V_{d - 2} \int_{r_h}^{\infty} \Lhd
  \mathcal{R}^{d - 2} \sqrt{| g_{t t} g_{r r} |} d r . \label{eqn:MLhdformula}
\end{equation}
Again, the simplicity of this answer suggests a more general principle at
work, but we defer further consideration of this to future work.

Unlike \eqref{eqn:SLhdformula}, it is not trivial to extend our derivation of
\eqref{eqn:MLhdformula} to Lagrangians $\Lhd$ involving
arbitrarily many covariant derivatives. Instead, we limit ourselves to a few observations. First, note that \eqref{eqn:MLhdformula} is correctly
unchanged by adding a total derivative to $\Lhd$, once
again because the lapse function vanishes at the horizon and the fields fall
off sufficiently rapidly at infinity. In appendix~\ref{sec.basis}, we show that
arbitrary three and four-derivative operators can be rewritten in terms of
$\nabla_{\mu} \phi^a$, $F_{\mu \nu}^A$, and $R^{\mu}_{\; \nu \rho
\sigma}$ via integration by parts, eliminating all higher covariant
derivatives. Thus, \eqref{eqn:MLhdformula} holds to \emph{at least} four-derivative
order, if not beyond.

\section{On the independence of mass and entropy corrections\label{sec.independence} }

At first glance, the mass and entropy corrections \eqref{eq.formulaM}, \eqref{eq.formulaS} appear to be related, especially when written in the form \eqref{eqn:simpMassFormula}, \eqref{eqn:simpEntropyFormula}. This may seem to confirm the claim~\cite{Cheung:2018cwt,Goon:2019faz} that they are directly (anti)correlated. However, notice that a naive reading of \eqref{eqn:simpMassFormula}, \eqref{eqn:simpEntropyFormula} suggests that $\delta M$ and $\delta \mathcal{S}$ should have the \emph{same} sign, whereas~\cite{Cheung:2018cwt,Goon:2019faz} argue that they have opposite signs.

\subsection{Demonstration of independence}   \label{sec.subindepedence}

In fact, despite appearances the extremal entropy correction \eqref{eq.formulaS}, \eqref{eqn:simpEntropyFormula} is \emph{independent} of the extremal mass correction \eqref{eq.formulaM}, \eqref{eqn:simpMassFormula}, in the sense that each one can have any magnitude or sign independent of the other in a generic effective field theory.\footnote{See~\cite{McPeak:2021tvu} for similar arguments in the special case of Reissner-Nordstr\"om black holes.}

To show this, it suffices to compare the effect of two different four-derivative operators:
\begin{align}
	\alpha' \Lhd^\text{example}= a_{abAB}(\phi) (F^A\cdot F^B)(\nabla \phi^a\cdot \nabla  \phi^b)+a_{ABCD}(\phi)(F^A\cdot F^B)(F^C\cdot F^D).
\end{align}
The resulting entropy correction is easily evaluated using \eqref{eqn:simpEntropyFormula}:
\be
\delta \mathcal{S} = - \frac{2\pi}{(d-3)^2 V_{d-2}^3 \mathcal{R}_h^{3d-8}} a^{A B C D}(\phi_h) Q_A Q_B Q_C Q_D \,,
\ee
where $\mathcal{R}_h = \mathcal{R}(r_h)$ is the curvature radius of the horizon, $\phi_h^a = \phi^a(r_h)$ is the attractor point in question, and $a^{A B C D}(\phi) \equiv f^{AA'}(\phi)\cdots f^{DD'}(\phi) a_{A'B'C'D'}(\phi)$. In particular, the $F^2 (\nabla\phi)^2$ coupling does not contribute to the extremal entropy correction because the moduli are constant in the near-horizon limit. On the other hand, the moduli are generically not constant far from the horizon, hence $F^2 (\nabla\phi)^2$ \emph{does} contribute to the extremal mass correction. Because of this, by adjusting the coefficient $a_{abAB}(\phi)$ we can choose the extremal mass correction to have any magnitude or sign, regardless of what the extremal entropy correction is.

While the above example demonstrates that the extremal mass and entropy corrections are independent, this independence is not limited to theories with moduli. For instance, consider the four-derivative operator
\be
\alpha' \Lhd^\text{example} = \hat{a}_{AB} (R^{\mu \nu \rho \sigma} - 2 R^{\mu \rho} g^{\nu \sigma}) F^A_{\mu \nu} F^B_{\rho \sigma} \,.
\ee
Due to the simplified form of the Riemann tensor in the near-horizon limit (see, e.g., appendix~\ref{app:Riemann}), this operator evaluates to zero in that limit and thus generates no entropy correction. However, it \emph{does} generically generate a mass correction, for instance
\be
\delta M = \frac{2 (d-3)^2}{(3d-7)V_{d-2} \mathcal{R}_h^{d-1}} \hat{a}^{A B} Q_A Q_B \,
\ee
in the Reissner-Nordstr\"om case, where $\hat{a}^{A B} \equiv f^{AA'} f^{BB'} \hat{a}_{A'B'}$ similar to above. This is made possible by the additional non-vanishing Riemann tensor components (mixing the angular and $t$--$r$ directions) that appear away from the near-horizon limit.

\subsection{Comparison with the literature} 

How can we reconcile this with the claim, due to~\cite{Cheung:2018cwt,Goon:2019faz}, that the entropy correction to a near-extremal black hole is positive if and only if the mass correction to the same black hole is negative?

The essential difference is that~\cite{Cheung:2018cwt,Goon:2019faz} consider the near-extremal entropy correction at \emph{fixed charge and fixed mass}, whereas our extremal calculations are at \emph{fixed charge and fixed (zero) temperature}. Before elaborating, we first reproduce the results of~\cite{Cheung:2018cwt,Goon:2019faz}. Since \eqref{eq.formulaM}, \eqref{eq.formulaS} apply only to extremal black holes, this requires some additional work.

We start by combining~\eqref{eq.psieqn} and~\eqref{eq.conseqn} to obtain:
\be
\frac d{dz}\Bigl[f\dot\psi-\frac{\chi}{2}\Bigr] -\frac{1}{2}\dot\psi\bigl[f\dot \psi+\dot f\bigr]=\frac{k_N}{2} f G_{a b} \dot{\phi}^a \dot{\phi}^b + \frac{k_N}{2} e^{2\psi}  Q^2(\phi) - \ap k_N e^{2 \psi} A^2 \biggl[ \Thd{}_t^t + F^A_{t r}\frac{\delta \Shd}{\delta F^A_{t r}} \biggr] . \label{eqn:Tttcor}
\ee
We expand the solution about an uncorrected solution,
\be
\psi =\psi_{(0)} + \delta \psi,\qquad
\phi^a = \phi^a_{(0)} + \delta \phi,\qquad
f = f_{(0)} + \delta\! f,
\ee
where $\delta \psi(z)$, $\delta \phi^a(z)$ and $\delta\! f(z)$ are $O(\ap)$ perturbations to the solution and we hold the charges $Q_A$ fixed. Substituting into~\eqref{eqn:Tttcor} and simplifying using the leading-order equations of motion, we obtain
\be
\frac{d}{dz} \biggl[ f \delta\dot{\psi} - \frac{\delta \chi}{2}  +\frac{\dot{\psi}}{2} \delta\! f - \frac{\dot{f}}{2} \delta\psi - f \dot{\psi} \delta \psi - k_N f G_{a b} \dot{\phi}^a \delta \phi^b \biggr] = - \ap k_N e^{2 \psi} A^2 \biggl[ \Thd{}_t^t + F^A_{t r}\frac{\delta \Shd}{\delta F^A_{t r}} \biggr] ,
\ee
up to $O(\ap^2)$, where we omit the $(0)$ subscripts on the leading-order solution for ease of notation.
Integrating $z$ from $0$ to $z_h$ (i.e., from $r=\infty$ to $r=r_h$) gives
\be
-\delta \dot{\psi}_\infty + \frac{\delta \chi_\infty}{2} -\frac{\delta \chi(z_h)}{2} 
  +\frac{\dot{\psi}(z_h)}{2} \delta\! f(z_h) + \frac{1}{2z_h} \delta\psi(z_h) = - \ap k_N\! \int_0^{z_h}\! e^{2 \psi} A^2 \biggl[ \Thd{}_t^t + F^A_{t r}\frac{\delta \Shd}{\delta F^A_{t r}} \biggr] d z , \label{eqn:prefirstlaw}
\ee
where we use $f_{(0)} = 1-\frac{z}{z_h}$ at leading order and $f_\infty = 1$, $\psi_\infty = 0$ to all orders, holding the asymptotic moduli values $\phi_\infty^a$ fixed.

In our chosen gauge, the coordinate location of the horizon $z_h$ is related to the surface gravity $g_h$ and horizon area $A_h$ via~\eqref{eqn:gAzRelation}. Thus, $z_h$ receives $\ap$ corrections, and we must carefully distinguish between, e.g., $\delta \psi(z_h)$, which is $\delta \psi(z)$ evaluated at the leading-order horizon $z=z_h^{(0)}$, versus the correction to the value of $\psi$ at the horizon, which is instead
\be
\delta \psi_h = \psi(z_h) - \psi^{(0)}(z_h^{(0)}) = \delta\psi(z_h) + \dot\psi(z_h) \delta z_h ,
\ee
up to terms that are $O(\alpha'^2)$.
Along similar lines, the gauge-fixing conditions~\eqref{eq.fBCs} ($f_h = 0$, $\dot{f}_h = -\frac{1}{z_h}$) and~\eqref{eq.chiBCs} ($\chi_h = \frac{1}{z_h}$, $\dot{\chi}_h =0$) imply that 
\begin{align}
\delta\! f(z_h) &= \frac{\delta z_h}{z_h} , & \delta \chi(z_h) &= -\frac{\delta z_h}{z_h^2} .
\end{align}
Thus,~\eqref{eqn:prefirstlaw} becomes:
\be
-\delta \dot{\psi}_\infty + \frac{\delta \chi_\infty}{2} 
   + \frac{1}{2z_h} \biggl[\delta\psi_h + \frac{\delta z_h}{z_h} \biggr]= - \ap k_N\! \int_0^{z_h}\! e^{2 \psi} A^2 \biggl[ \Thd{}_t^t + F^A_{t r}\frac{\delta \Shd}{\delta F^A_{t r}} \biggr] d z ,
\ee
up to $O(\ap^2)$. Using~\eqref{eq.Ahofxh}, \eqref{eqn:gAzRelation}, and \eqref{eq.ADMmass}, this can be rewritten as
\be
\delta M - \frac{1}{\kappa^2} g_h \delta A_h = - \ap \int_0^{z_h}\! e^{2 \psi} A^2 \biggl[ \Thd{}_t^t + F^A_{t r}\frac{\delta \Shd}{\delta F^A_{t r}} \biggr] d z \,.
\ee
Finally, using~\eqref{eq.SSplit} to relate the change in area to the change in entropy and rearranging, we find
\begin{multline}
\biggl[\delta M - \frac{g_h}{2\pi} \delta\mathcal{S} \biggr]_{\text{fixed $Q,\phi_\infty^a$}} = - \ap V_{d-2}  \int_{r_h}^\infty \biggl(\Thd{}^t_t + F_{t r}^A \frac{\delta \Shd}{\delta F^A_{tr}} \biggr) \mathcal{R}^{d-2} \sqrt{|g_{t t} g_{r r}|} dr \\ - \ap g_h A_h \frac{\delta \Shd}{\delta R^{tr}\!_{tr}}\biggr|_{r=r_h}+ O(\ap^2), \label{eqn:deltaMSreln}
\end{multline}
where $g_h$ is the surface gravity, related to the Hawking temperature $T_{\rm BH} = \frac{g_h}{2\pi}$. Note that the left-hand-side of~\eqref{eqn:deltaMSreln} resembles the first law of black hole mechanics, but is technically distinct from it since we are computing the change in the solution induced by the $\ap$ corrections, rather than varying the solution with fixed $\ap$ corrections.

\subsubsection*{Entropy corrections at fixed mass versus fixed temperature}

Using~\eqref{eqn:deltaMSreln}, we can deduce several things. Firstly, in the zero temperature limit $g_h \to 0$ we recover the extremal (i.e., fixed charge and fixed zero temperature) mass correction~\eqref{eq.formulaM}. Alternately, per~\eqref{eqn:deltaMSreln}, the mass correction at fixed charge and \emph{fixed entropy} is given by
\begin{multline}
\delta M \biggr|_{\text{fixed $Q,\phi_\infty^a,\mathcal{S}$}} = - \ap V_{d-2}  \int_{r_h}^\infty \biggl(\Thd{}^t_t + F_{t r}^A \frac{\delta \Shd}{\delta F^A_{tr}} \biggr) \mathcal{R}^{d-2} \sqrt{|g_{t t} g_{r r}|} dr \\ - \ap g_h A_h \frac{\delta \Shd}{\delta R^{tr}\!_{tr}}\biggr|_{r=r_h}+ O(\ap^2). \label{eq.dMfixedS}
\end{multline}
This once again reduces to the extremal mass correction~\eqref{eq.formulaM} in the zero temperature limit $g_h \to 0$, but for a subtle reason: although fixed temperature and fixed entropy are \emph{not} the same in general---in particular, the extremal entropy correction (at fixed, zero temperature) is in general nonzero---the mass correction becomes insensitive to the difference in the zero temperature limit because of the $g_h$ in front of $\delta \mathcal{S}$ in~\eqref{eqn:deltaMSreln}. 

On the other hand, \eqref{eqn:deltaMSreln} also implies that the \emph{entropy} correction at fixed charge and \emph{fixed mass} is given by
\begin{multline}
\delta \mathcal{S} \biggr|_{\text{fixed $Q,\phi_\infty^a,M$}} = \frac{2 \pi \ap}{g_h} V_{d-2}  \int_{r_h}^\infty \biggl(\Thd{}^t_t + F_{t r}^A \frac{\delta \Shd}{\delta F^A_{tr}} \biggr) \mathcal{R}^{d-2} \sqrt{|g_{t t} g_{r r}|} dr \\ + 2 \pi \ap A_h \frac{\delta \Shd}{\delta R^{tr}\!_{tr}}\biggr|_{r=r_h}+ O(\ap^2). \label{eq.dSfixedM}
\end{multline}
This \emph{does not} reduce to the extremal entropy correction~\eqref{eq.formulaS} in the zero temperature limit; in particular, the first term diverges in this limit, whereas \eqref{eq.formulaS} is finite. The reason is simply that fixed mass and fixed temperature are generally distinct---the extremal mass correction being generally nonzero---whereas the same factor of $g_h$ in front of $\delta \mathcal{S}$ in~\eqref{eqn:deltaMSreln} makes the entropy correction \emph{hypersensitive} to the difference in the $g_h \to 0$ limit.\footnote{How can a finite, extremal entropy correction emerge from~\eqref{eqn:deltaMSreln}? Working at \emph{fixed temperature},
\begin{multline}
\delta\mathcal{S} \biggr|_{\text{fixed $Q,\phi_\infty^a, T$}} = \frac{2 \pi}{g_h} \biggl[\delta M \biggr|_{\text{fixed $Q,\phi_\infty^a, T$}}+ \ap V_{d-2}  \int_{r_h}^\infty\!\! \biggl(\Thd{}^t_t + F_{t r}^A \frac{\delta \Shd}{\delta F^A_{tr}} \biggr) \mathcal{R}^{d-2} \sqrt{|g_{t t} g_{r r}|} dr\biggr] + \\ + 2 \pi \ap A_h \frac{\delta \Shd}{\delta R^{tr}\!_{tr}}\biggr|_{r=r_h}+ O(\ap^2).
\end{multline}
Expanding about zero temperature and comparing with the extremal (fixed temperature) mass correction formula~\eqref{eq.formulaM}, we see that the term in brackets is $O(g_h)$, avoiding a divergence. However, to actually reproduce the extremal entropy formula~\eqref{eq.formulaS} we would need to calculate the near-extremal mass correction to $O(g_h)$, which is beyond the scope of this paper.
}

In summary, the entropy correction near extremality depends sensitively on whether we hold the mass or the temperature fixed (along with the charge). In our work, we computed the correction~\eqref{eq.formulaS} to the entropy of an extremal black, holding the temperature fixed (at zero). As shown in~\S\ref{sec.subindepedence}, the extremal mass and entropy corrections are independent, and their signs can be the same or different depending on the choice of effective field theory.

By contrast, \cite{Cheung:2018cwt,Goon:2019faz} consider the near extremal entropy correction at fixed charge \emph{and fixed mass}. Then, comparing~\eqref{eq.dMfixedS} and~\eqref{eq.dSfixedM}, one concludes that
\be
\delta M\biggr|_{\text{fixed $Q,\phi_\infty^a,\mathcal{S}$}} = -\frac{g_h}{2\pi} \delta\mathcal{S} \biggr|_{\text{fixed $Q,\phi_\infty^a,M$}} , \label{eqn:dMdSreln}
\ee
as first shown in \cite{Cheung:2018cwt,Goon:2019faz}. Thus, the extremal mass correction (being insensitive to the distinction between fixed temperature and fixed entropy) is negative if and only if the near-extremal entropy correction \emph{at fixed mass} (and charge) is positive.

Thus, our results do not disagree with those of~\cite{Cheung:2018cwt,Goon:2019faz}. A deeper question that we will not attempt to answer is which notion of entropy correction is relevant in various contexts. Arguably, the extremal entropy correction that we have calculated is a more ``natural'' quantity than the near-extremal, fixed-mass entropy correction that appears in~\eqref{eqn:dMdSreln}, for instance because the former is finite whereas the latter diverges at zero temperature. However, when arguing that $\delta \mathcal{S} > 0$ (as in \cite{Cheung:2018cwt}) either (or neither) notion might be the correct, depending on the argument. Here we simply emphasize the difference without addressing these deeper questions.

\section{Examples \label{sec.examples}}

We now consider a few explicit examples to further illustrate our methods.

\subsection{Electric Reissner-Nordstr{\"o}m black holes} \label{sec.eRN}

We begin with the simplest case of Einstein-Maxwell theory, with the two-derivative effective action:
\begin{equation}
  S = \int d^d x \sqrt{- g}  \biggl( \frac{1}{2 \kappa_d^2} R - \frac{1}{2
  e_d^2} F^2 \biggr),
\end{equation}
where $\kappa_d$ and $e_d$ are the gravitational and gauge couplings of
dimensions $- \frac{d - 2}{2}$ and $- \frac{d - 4}{2}$, respectively. The
(Reissner-Nordstr{\"o}m) extremal charged black hole solutions are most
conveniently expressed in the gauge\footnote{Note that, although this differs
from the gauge introduced in \S\ref{sec.eeom}, since the formulas~\eqref{eqn:basicResults}, \eqref{eqn:simpFormulas} are invariant under radial gauge changes we can use any
convenient gauge.}
\begin{align}
  d s^2 & = - \Biggl[ 1 - \frac{\mathcal{R}_h^{d - 3}}{r^{d - 3}} \Biggr]
  d t^2 + \Biggl[ 1 - \frac{\mathcal{R}_h^{d - 3}}{r^{d - 3}} \Biggr]^{-
  1} \!\!\! d r^2 + r^2 d \Omega_{d - 2}^2, \qquad \mathcal{R}_h^{d - 3}
  \equiv \frac{\sqrt{k_N} e_d | Q |}{(d - 3) V_{d - 2}}, \nonumber\\
  F & = - \frac{e^2_d Q}{V_{d - 2} r^{d - 2}} d t \wedge d r, 
\end{align}
with mass $M_0 = \frac{e_d | Q |}{\sqrt{k_N}}$, where $k_N = \frac{d - 3}{d - 2} \kappa_d^2$ as before.

Per the results of \S\ref{subsec:parityeven}, all possible parity-even
four-derivative corrections to this theory can be reduced to four independent
couplings:\footnote{Recall that $F^2 \equiv \frac{1}{2} F_{\mu \nu} F^{\mu
\nu}$ in our conventions.}
\begin{equation}
  \mathcal{L}_{(4)} = c_{\text{GB}} \mathcal{L}_{\text{GB}} + c_{R F^2} R^{\mu
  \nu \rho \sigma} F_{\mu \nu} F_{\rho \sigma} + c_{(F^2)^2}  (F^2)^2 +
  c_{F^4} F_{\mu \nu} F^{\nu \rho} F_{\rho \sigma} F^{\sigma \mu},
  \label{eqn:RNL4}
\end{equation}
where $\mathcal{L}_{\text{GB}} = R^{\mu \nu \rho \sigma} R_{\mu \nu \rho
\sigma} - 4 R^{\mu \nu} R_{\mu \nu} + R^2$ is the Gauss-Bonnet combination.
Applying \eqref{eqn:simpFormulas}, one finds the mass and entropy corrections
\begin{subequations} \label{eqn:RNeMScorrs}
\begin{align} 
  \delta M & = (d \!-\! 3) V_{d - 2} \mathcal{R}_h^{d - 5} \! \biggl[ (d\! -\! 2) (d\! -\! 4)
  c_{\text{GB}} - \frac{(d\! -\! 3)^3}{3 d\! -\! 7}  \biggl( \frac{2 e_d^2 c_{R
  F^2}}{k_N} + \frac{e^4_d [c_{(F^2)^2} \!+\! 2 c_{F^4}]}{k_N^2} \biggr)\! \biggr]\!, 
  \label{eqn:RNeMcorr}\\
  \delta \mathcal{S} & = 2 \pi V_{d - 2} \mathcal{R}_h^{d - 4} \! \biggl[ \frac{(d \!
  - \!2) [3 d^2 \!-\! 15 d \!+\! 16]}{d \!-\! 3} c_{\text{GB}} - (d \!-\! 3)^2 \! \biggl( \frac{4
  e^2_d c_{R F^2}}{k_N} + \frac{e^4_d [c_{(F^2)^2} \!+\! 2 c_{F^4}]}{k_N^2} 
  \biggr) \! \biggr] \!,\label{eqn:RNeScorr}
\end{align}
\end{subequations}
where $\mathcal{R}_h^{d-3} = \frac{\sqrt{k_N} e_d | Q |}{(d - 3) V_{d - 2}} = \frac{k_N M_0}{(d - 3) V_{d - 2}}$.

A few comments are in order. First, note that \eqref{eqn:RNeMcorr} reproduces
the results of \cite{Kats:2006xp}, appendix B. Second, we observe that both the mass and
entropy corrections depend on $c_{(F^2)^2}$ and $c_{F^4}$ in the combination
$c_{(F^2)^2} + 2 c_{F^4}$. This is a consequence of parity and spherical
symmetry, as explained in \S\ref{subsec:spherical}.

On the other hand, the remaining couplings all appear in independent ways in
the mass and entropy corrections, demonstrating the general results of
\S\ref{sec.independence}. To illustrate this, consider the 4d case:
\begin{subequations} \label{eqn:RNeCorr4d}
\begin{align}
  \delta M^{(d = 4)} & = - \frac{16 \pi^2}{5 k_N M_0}  \left( \frac{2 e_4^2
  c_{R F^2}}{k_N} + \frac{e^4_4 [c_{(F^2)^2} + 2 c_{F^4}]}{k_N^2} \right),
  \\
  \delta \mathcal{S}^{(d = 4)} & = 8 \pi^2 \left[ 8 c_{\text{GB}} - \frac{4
  e^2_4 c_{R F^2}}{k_N} - \frac{e^4_4 [c_{(F^2)^2} + 2 c_{F^4}]}{k_N^2}
  \right] .  
\end{align}
\end{subequations}
In particular, notice that the Gauss-Bonnet operator contributes to the
entropy correction but not the mass correction. This operator is actually
topological (locally a total derivative) in 4d, explaining its vanishing
contribution to the mass correction, which depends only on the equations of
motion. On the other hand, higher-derivative topological operators
\emph{can} correct the (Wald) entropy~\cite{Jacobson:1993xs,Liko:2007vi,Sarkar:2010xp,Bobev:2021oku}, as happens here. However, since the operators $R^{\mu \nu \rho
\sigma} F_{\mu \nu} F_{\rho \sigma}$ and $(F^2)^2$ \emph{also} contribute to the mass
and entropy in linearly independent ways, the independence of the mass and
entropy corrections does not rely on this subtle point about topological
operators.

With a little more effort (see the helpful formulas in appendix
\ref{app:stresstensor}), one can reproduce \eqref{eqn:RNeMScorrs}
using \eqref{eqn:basicResults}, from which we also obtain the self-force correction
\begin{equation}
  \hat{F}_{\text{self}} = - 2 (d \!-\! 3)^2 V_{d - 2}^2 \mathcal{R}_h^{2 (d - 4)} 
  \biggl[ (d \!-\! 2) (d \!-\! 4) c_{\text{GB}} - \frac{(d \!-\! 3)^3}{3 d \!-\! 7}  \biggl(
  \frac{2 e_d^2 c_{R F^2}}{k_N} + \frac{e^4_d [c_{(F^2)^2} \!+\! 2
  c_{F^4}]}{k_N^2} \biggr)\! \biggr]\! . \label{eqn:RNeFcorr}
\end{equation}
In fact, the mass and force corrections are not independent in the absence of
moduli, since $\hat{F}_{\text{self}} = e_d^2 Q^2 - k_N M^2 = - 2 k_N M_0
\delta M + O (\delta M^2)$ upon substituting in the corrected mass $M = M_0 +
\delta M$. This relation indeed holds for \eqref{eqn:RNeMcorr},
\eqref{eqn:RNeFcorr}.

\subsection{Dyonic Reissner-Nordstr{\"o}m black holes} \label{sec.dRN}

We now turn to 4d dyonic Reissner-Nordstr{\"o}m black holes, in part as a
natural extension of the above and in part as a preview of the dyonic
Einstein-Maxwell dilaton black holes to be discussed below. The leading-order
solution is now
\begin{align}
  d s^2 &= - \left[ 1 - \frac{\mathcal{R}_h}{r} \right] \! d t^2 +
  \left[ 1 - \frac{\mathcal{R}_h}{r} \right]^{- 1} \!\!\! d r^2 + r^2 d
  \Omega_2^2, & \mathcal{R}_h &\equiv \frac{\sqrt{k_N (e^2 Q_e^2 +
  \tilde{e}^2 Q_m^2)}}{4 \pi} , \nonumber\\
  F &= - \frac{e^2 Q_e}{4 \pi r^2} d t \wedge d r + \frac{Q_m \sin
  \theta}{2} d \theta \wedge d \varphi, 
\end{align}
with mass-squared $M_0^2 = \frac{e^2 Q_e^2 +
  \tilde{e}^2 Q_m^2}{k_N}$,
where for simplicity we set the theta angle to zero and $\tilde{e} \equiv 2
\pi / e$ is the magnetic gauge coupling. Defining $\zeta \equiv \bigl|
\frac{\tilde{e} Q_m}{e Q_e} \bigr|$, we obtain
\begin{subequations} \label{eqn:RNdCorrs}
\begin{align}
  \delta M & = - \frac{16 \pi^2}{5 k_N M_0}  \left[ \frac{1 + 3 \zeta^2}{1 +
  \zeta^2}  \frac{2 e^2 c_{R F^2}}{k_N} + \frac{(1 - \zeta^2)^2}{(1 +
  \zeta^2)^2}  \frac{e^4 c_{(F^2)^2}}{k_N^2} + \frac{2 (1 + \zeta^4)}{(1 +
  \zeta^2)^2}  \frac{e^4 c_{F^4}}{k_N^2} \right], \\
  \delta \mathcal{S} & = 8 \pi^2  \left[ 8 c_{\text{GB}} - \frac{4 e^2 c_{R
  F^2}}{k_N} - \left[ \frac{1 - \zeta^2}{1 + \zeta^2} \right]^2  \frac{e^4
  c_{(F^2)^2}}{k_N^2} - \frac{2 (1 + \zeta^4)}{[1 + \zeta^2]^2}  \frac{e^4
  c_{F^4}}{k_N^2} \right] ,
\end{align}
\end{subequations}
assuming the same four-derivative operators \eqref{eqn:RNL4} are present. The
self-force coefficient can likewise be computed, and comes out to
$\hat{F}_{\text{self}} = - 2 k_N M_0 \delta M$ as expected.

Note that in principle the results \eqref{eqn:RNdCorrs} can be deduced from
\eqref{eqn:RNeCorr4d} using electromagnetic duality, though doing so is not
completely straightforward. To illustrate this, we consider the
effect of S-duality, $Q_e' = Q_m, Q_m' = - Q_e$, $e' = \tilde{e}$, and $F' =
\frac{2 \pi}{e^2}  \tilde{F}$ where $\tilde{F} \equiv - \star F$. This takes
$\zeta \rightarrow 1 / \zeta$, but also changes the coefficients of the
higher-derivative operators in \eqref{eqn:RNL4}. In particular
\begin{align}
  \mathcal{L}_{(4)}' &= c_{\text{GB}}' \mathcal{L}_{\text{GB}}' + c_{R F^2}'
  R^{\mu \nu \rho \sigma} F_{\mu \nu}' F_{\rho \sigma}' + c_{(F^2)^2}'  (
  {F'}^2 )^2 + c_{F^4}' F_{\mu \nu}' {F^{\nu \rho}}' F_{\rho \sigma}'
  {F^{\sigma \mu}}' \nonumber\\
  &= c_{\text{GB}}' \mathcal{L}_{\text{GB}} + \biggl[ \frac{e'}{e } \biggr]^2\!
  c_{R F^2}' R^{\mu \nu \rho \sigma} \tilde{F}_{\mu \nu} \tilde{F}_{\rho
  \sigma} + \biggl[ \frac{e'}{e } \biggr]^4 \! c_{(F^2)^2}'  (\tilde{F}^2)^2 +
  \biggl[ \frac{e'}{e } \biggr]^4 \! c_{F^4}'  \tilde{F}_{\mu \nu} \tilde{F}^{\nu
  \rho} \tilde{F}_{\rho \sigma} \tilde{F}^{\sigma \mu} . 
\end{align}
Eliminating pairs of $\tilde{F}$'s using $\Omega_{\mu \nu \rho \sigma}
\Omega^{\alpha \beta \gamma \delta} = - 24 \delta_{[\mu }^{\alpha}
\delta_{\nu}^{\beta} \delta_{\rho}^{\gamma} \delta_{
\sigma]}^{\delta}$, we obtain
\begin{multline}
  \mathcal{L}_{(4)}' = c_{\text{GB}}' \mathcal{L}_{\text{GB}} - \biggl[
  \frac{e'}{e } \biggr]^2 c_{R F^2}'  (R^{\mu \nu \rho \sigma} F_{\mu \nu}
  F_{\rho \sigma} - 4 R^{\mu}_{\nu} F^{\nu \rho} F_{\mu \rho} + R F^{\mu \nu}
  F_{\mu \nu}) \\ + \biggl[ \frac{e'}{e } \biggr]^4 c_{(F^2)^2}'  (F^2)^2
  + \biggl[ \frac{e'}{e } \biggr]^4 c_{F^4}' F_{\mu \nu} F^{\nu \rho} F_{\rho
  \sigma} F^{\sigma \mu} . 
\end{multline}
Next, we use the leading-order
Einstein equations $R_{\mu \nu} = \frac{\kappa_4^2}{e^2}  \left( F_{\mu} \cdot
F_{\nu} - \frac{1}{2} g_{\mu \nu} F^2 \right)$ to put this back into the form \eqref{eqn:RNL4}, 
\begin{multline}
  \mathcal{L}_{(4)}' = c_{\text{GB}}' \mathcal{L}_{\text{GB}} - \biggl[
  \frac{e'}{e } \biggr]^2 c_{R F^2}' R^{\mu \nu \rho \sigma} F_{\mu \nu}
  F_{\rho \sigma} + 8 k_N  \biggl[ \frac{e'}{{e^2} } \biggr]^2 c_{R F^2}' 
  (F_{\mu \nu} F^{\nu \rho} F_{\rho \sigma} F^{\sigma \mu} - (F^2)^2)
  \\
  + \biggl[ \frac{e'}{e } \biggr]^4 c_{(F^2)^2}'  (F^2)^2 + \biggl[ \frac{e'}{e
  } \biggr]^4 c_{F^4}' F_{\mu \nu} F^{\nu \rho} F_{\rho \sigma} F^{\sigma \mu}
  ,
\end{multline}
from which we read off
\begin{equation}
\begin{aligned}
  c_{\text{GB}} &= c_{\text{GB}}', &  e^4 c_{(F^2)^2} &= e'^4 c_{(F^2)^2}' - 8 k_N {e'}^2 c_{R F^2}', \\
  e^2 c_{R F^2} &= - e'^2 c_{R F^2}',  &
  e^4 c_{F^4} &= e'^4 c_{F^4}' + 8 k_N {e'}^2 c_{R F^2}' .
 \end{aligned}
\end{equation}
One can check that, together with $\zeta' = 1 / \zeta$, this transformation
leaves \eqref{eqn:RNdCorrs} unchanged as required.

Thus, using S-duality we can deduce the purely magnetic $\zeta \rightarrow
\infty$ limit of \eqref{eqn:RNdCorrs} from the purely electric result
\eqref{eqn:RNeCorr4d}. However, deriving \eqref{eqn:RNdCorrs} in its entirety
from \eqref{eqn:RNeCorr4d} requires a more general calculation (e.g., using a
democratic approach), which we omit for the sake of brevity.

Let us examine the special case $\zeta = 1$ more closely (for which the
electric and magnetic fields have equal magnitude):
\begin{align}
  \delta M^{(\zeta = 1)} &= - \frac{16 \pi^2}{5 k_N M_0}  \biggl[ \frac{4 e^2
  c_{R F^2}}{k_N} + \frac{e^4 c_{F^4}}{k_N^2} \biggr], & \delta
  \mathcal{S}^{(\zeta = 1)} &= 8 \pi^2  \biggl[ 8 c_{\text{GB}} - \frac{4 e^2
  c_{R F^2}}{k_N} - \frac{e^4 c_{F^4}}{k_N^2} \biggr] .
  \label{eqn:RNequaldyon}
\end{align}
We have repeatedly made the point that the extremal mass and entropy
corrections are independent, and that this independence does not depend on
topological couplings such as the 4d Gauss-Bonnet term. Nonetheless, if we
ignore the Gauss-Bonnet contribution then the linear relation $\delta M =
\frac{2}{5 k_N M_0} \delta \mathcal{S}$ seems to hold. What is going on here?

The answer is that we have set the parity-odd higher-derivative couplings to
zero for simplicity, even though the background we are studying is not parity
invariant. Per the analysis of appendix \ref{sec.basis}, there are two
additional parity-odd couplings that we should consider, $R^{\mu \nu \rho
\sigma} F_{\mu \nu} \tilde{F}_{\rho \sigma}$ and $\tilde{F}_{\mu \nu} F^{\nu
\rho} F_{\rho \sigma} F^{\sigma \mu}$. The latter vanishes for $\zeta = 1$,
and thus does not contribute to either the mass or the entropy corrections.
The former \emph{does} contribute, but only to the mass:
\begin{align}
  \delta M_{R F \tilde{F}}^{(\zeta = 1)} &= \pm \frac{32 \pi^2 e^2 c_{R F
  \tilde{F}}}{5 k_N^2 M_0}, & \delta \mathcal{S}_{R F \tilde{F}}^{(\zeta
  = 1)} &= 0,
\end{align}
where the overall sign is that in $e Q_e = \pm \tilde{e} Q_m$. Thus,
upon turning on all possible couplings, the independence of the mass and
entropy corrections is again manifest.

Finally, note that the WGC constraint $\delta M \leqslant 0$ is more powerful
when applied to the full spectrum of dyonic black holes, rather than just
electrically-charged black holes. In particular, define the dimensionless
combinations
\begin{equation}
  c_1 \equiv \frac{e^2 c_{R F^2}}{k_N}, \qquad c_2 \equiv \frac{e^4
  c_{(F^2)^2}}{k_N^2} - 4 \frac{e^2 c_{R F^2}}{k_N}, \qquad c_3 \equiv
  \frac{e^4 c_{F^4}}{k_N^2} + 4 \frac{e^2 c_{R F^2}}{k_N} .
\end{equation}
Then, in terms of $u = 2 \log \zeta$ one finds the mass correction
\begin{equation}
  \delta M = - \frac{16 \pi^2}{5 k_N M_0}  \frac{[(c_2 + 2 c_3) \cosh u + 2
  c_1 \sinh u - c_2]}{\cosh u + 1},
\end{equation}
in the absence of parity-odd couplings. This is negative semi-definite for all
$u$ iff
\begin{equation}
  c_2 + 2 c_3 \geqslant 2 | c_1 | \qquad \text{and} \qquad \sqrt{(c_2 + 2
  c_3)^2 - 4 c_1^2} \geqslant c_2 . \label{eqn:dyonicWGCcons}
\end{equation}
By comparison, only considering the electric case ($u = - \infty$) yields the
weaker constraint $c_2 + 2 c_3 \geqslant 2 c_1$.

These constraints will change when we include parity-odd operators. However,
since parity-odd contributions are always odd under $Q_m \rightarrow - Q_m$
with $Q_e$ fixed (leaving $u = \log \frac{\tilde{e}^2 Q_m^2}{e^2 Q_e^2}$ also
fixed) the WGC bound $\delta M \leqslant 0$ only gets harder to satisfy, and
\eqref{eqn:dyonicWGCcons} is still a necessary condition.

\subsection{Dyonic Einstein-Maxwell-Dilaton black holes} \label{sec.EMdexample}

We now generalize our discussion to the case with moduli. Perhaps the simplest
two-derivative effective field theory involving a modulus coupled a gauge
field and gravity is Einstein-Maxwell-Dilaton theory, with the action:
\begin{equation}
  S = \int d^d x \sqrt{- g}  \left[ \frac{1}{2 \kappa^2}  \left( R -
  \frac{1}{2 \alpha^2}  (\nabla \phi)^2 \right) - \frac{1}{2 e^2_0} e^{- \phi}
  F \cdot F \right],
\end{equation}
where $\phi$ is the dilaton, $\alpha > 0$ is its dimensionless coupling
strength, and we set $\langle \phi \rangle = 0$ in the asymptotic vacuum by
convention.

Now, however, there are two difficulties. Firstly, the electrically charged
extremal black hole solutions in this theory have vanishing horizon area,
hence the derivative expansion breaks down near the horizon and we cannot
compute the corrections to their mass, entropy and self-force in effective
field theory. To overcome this difficulty, we consider 4d dyonic black holes,
for which the charge function
\begin{equation}
  Q^2 (\phi) = e^{\phi} e^2_0 Q_e^2 + e^{- \phi}  \tilde{e}^2_0 Q_m^2,
\end{equation}
has a minimum at the attractor point $\phi_h = \log \zeta_0$ for $\zeta_0
\equiv \left| \frac{\tilde{e}_0 Q_m}{e_0 Q_e} \right|$. Then, since $Q^2
(\phi_h) = 4 \pi | Q_e Q_m | > 0$, the horizon area is non-zero.

The second difficulty is more technical: while numerically tractable, these
dyonic solutions cannot be written in closed form except for the special cases
$\alpha = 0, 1, \sqrt{3}$. Note that $\alpha = 0$ is the
Reissner-Nordstr{\"o}m case, whereas $\alpha = \sqrt{3}$ arises naturally in
Kaluza-Klein theory. We instead focus on $\alpha = 1$, which arises naturally
in string theory. The extremal solution is then
\begin{align}
  d s^2 & = - e^{2 \psi} d t^2 + e^{- 2 \psi} [d r^2 + r^2
  d \Omega_2^2], \nonumber\\
  F & = - \frac{e_0^2 Q_e e^{2 \psi + \phi}}{4 \pi r^2} d t \wedge d
  r + \frac{Q_m \sin \theta}{2} d \theta \wedge d \varphi,
  \nonumber\\
  \psi \pm \frac{\phi}{2} & = - \log \biggl[ 1 + \frac{\mathcal{R}_{\pm}}{r}
  \biggr] \qquad \text{where} \qquad \mathcal{R}_{\pm} \equiv \frac{\sqrt{2 k_N}}{4 \pi}
\begin{cases}
e_0 |Q_e|, & +, \\
\tilde{e}_0 |Q_m|, & -,
\end{cases}
\end{align}
with mass
  $M_0 = \frac{| e_0 Q_e | + | \tilde{e}_0 Q_m |}{\sqrt{2 k_N}}$.

Imposing parity for simplicity, the possible four-derivative operators take
the form
\begin{align}
  \mathcal{L}_{(4)} & = a_{\text{GB}} (\phi) R_{\text{GB}} + a_{R F^2} (\phi)
  R^{\mu \nu \rho \sigma} F_{\mu \nu} F_{\rho \sigma} + a_{(F^2)^2} (\phi)
  (F^2)^2 + a_{F^4} (\phi) F_{\mu \nu} F^{\nu \rho} F_{\rho \sigma} F^{\sigma
  \mu} \nonumber\\
  & \noeq + a_{F^2 (\nabla \phi)^2} (\phi) F^2 (\nabla \phi)^2 + a_{(F
  \nabla \phi)^2} (\phi) F^{\mu \nu} F_{\mu \rho} \nabla_{\nu} \phi
  \nabla^{\rho} \phi + a_{(\nabla \phi)^4} (\phi) (\nabla \phi)^4, 
  \label{eqn:genericL4}
\end{align}
where $a_{\text{GB}} (\phi)$, $a_{R F^2} (\phi)$, etc., are a priori unknown
functions of the moduli. The entropy correction is easily evaluated using
\eqref{eqn:simpEntropyFormula}:
\begin{equation}
  \delta \mathcal{S}= 8 \pi^2  \left( 8 a_{\text{GB}} (\phi_h) - \frac{4 e^2
  (\phi_h) a_{R F^2} (\phi_h)}{k_N} - \frac{e^4_h (\phi_h) a_{F^4}
  (\phi_h)}{k_N^2} \right), \label{eqn:EMDgeneralS}
\end{equation}
where $e^2 (\phi) \equiv e_0^2 e^{\phi}$ is the dilaton-dependent gauge
coupling and $\phi_h = \log \zeta_0 = \log \left| \frac{\tilde{e}_0 Q_m}{e_0
Q_e} \right|$ is the attractor point. Note the strong similarity with
\eqref{eqn:RNequaldyon}. Indeed,
\begin{equation}
  \zeta (\phi_h) = \left| \frac{\tilde{e} (\phi_h) Q_m}{e (\phi_h) Q_e}
  \right| = 1,
\end{equation}
so the attractor mechanism automatically makes the electric and magnetic
fields equal in magnitude at the horizon, explaining why the entropy correction closely
parallels that of the $\zeta = 1$ dyonic Reissner-Nordstr{\"o}m case discussed
above.

On the other hand, to compute the mass correction we need to do a non-trivial
integral that depends on the functional form of the EFT coefficients in
(\ref{eqn:genericL4}). For example, in the case of the $F_{\mu \nu} F^{\nu
\rho} F_{\rho \sigma} F^{\sigma \mu}$ correction this integral can be written
as
\begin{equation}
  \delta M = - \frac{4 \pi^2 e_0^4}{k_N^3 M_0}  \int_0^{\phi_h} \frac{e^{- 2
  \phi}  [e^{\phi_h} + 1]  (e^{\phi} - 1)^4 (e^{4 \phi} + e^{4
  \phi_h})}{(e^{\phi_h} - 1)^5} a_{F^4} (\phi) d \phi .
\end{equation}
Similar expressions (of varying complexity) can be written for the other
operators in \eqref{eqn:genericL4}.

To obtain a more explicit result, we specialize to the four-derivative
Lagrangian
\begin{equation}
  \mathcal{L}_{(4)} = c_{\text{GB}} e^{- \phi}
  \mathcal{L}_{\text{GB}} + c_{(F^2)^2} e^{- 3 \phi}  (F^2)^2 + c_{F^4} e^{- 3
  \phi} F_{\mu \nu} F^{\nu \rho} F_{\rho \sigma} F^{\sigma \mu} + c_{F^2
  (\nabla \phi)^2} e^{- 2 \phi} F^2 (\nabla \phi)^2 . \label{eqn:HetL4}
\end{equation}
Here we have kept only certain terms in \eqref{eqn:genericL4} for simplicity,
and we have assumed a particular $\phi$ dependence, with the following
rationale (as in, e.g., \cite{Loges:2019jzs}). Suppose we begin with a four-dimensional
``string-frame'' action of the form
\begin{equation}
  S_{\text{str}} = \frac{1}{2 \kappa^2} \int d^4 x \sqrt{- g} e^{- 2 \Phi} 
  \left[ R + 4 (\nabla \Phi)^2 - \frac{\kappa^2}{e_0^2} F \cdot F +
  c_{\text{GB}} \mathcal{L}_{\text{GB}} + \cdots \right] ,
\end{equation}
where the overall factor of $e^{-2\Phi}$ occurs for closed strings at string tree-level.
Switching to Einstein frame:
\begin{equation}
  S_{\text{Ein}} = \frac{1}{2 \kappa^2} \int d^4 x \sqrt{- g}  \left[ R - 2
  (\nabla \Phi)^2 - \frac{\kappa^2}{e_0^2} e^{- 2 \Phi} F \cdot F +
  c_{\text{GB}} e^{- 2 \Phi} \mathcal{L}_{\text{GB}} + \cdots \right].
\end{equation}
Identifying $\phi = 2 \Phi$, we reproduce the
$\phi$-dependence seen in each term of \eqref{eqn:HetL4}.

To state the resulting corrections more concisely, it is convenient to define\footnote{Alternately, $f_p (\zeta)
= p \Phi (1 - \zeta, 1, p)$ in terms of the Lerch trancendent $\Phi (z, s, a)$.}
\begin{equation}
  f_p (\zeta_0) \equiv - p \frac{\log (\zeta_0) + \sum_{n = 1}^{p - 1}
  \frac{(1 - \zeta_0)^n}{n}}{(1 - \zeta_0)^p},
\end{equation}
for any positive integer $p$. This combination is chosen so that $f_p (1) =
1$, cancelling the apparent pole at $\zeta_0 = 1$. The mass correction is then\footnote{In comparison with~\cite{Loges:2019jzs}, our $(F^2)^2$ and Gauss-Bonnet corrections agree, but we obtain the opposite sign for the $F^2 (\nabla \phi)^2$ correction. The basis used in \cite{Loges:2019jzs} does not include an $F^4$ term. While this can be related to the $(F \cdot \tilde{F})^2$ term that they do include, they implicitly choose a different dilaton coupling for this term, preventing a direct comparison of our $F^4$ correction with their results.}
\begin{multline}
  \delta M  = - \frac{2 \pi^2}{5 k_N M}  \frac{1 +
  \zeta_0}{\zeta_0} \biggl[ 8(2 + 5 f_1 - 20 f_2 + 20 f_3 - 10 f_4 + 3 f_5)
  c_{\text{GB}} \\
   \qquad \qquad \quad +  (1 -
  20 f_2 + 40 f_3 - 25 f_4 + 4 f_5)  \frac{e_0^4 c_{(F^2)^2}}{k_N^2} 
   +2(1 + 5 f_4 - 4 f_5) 
  \frac{e_0^4 c_{F^4}}{k_N^2} \\
   +2(1 -
  10 f_1 + 20 f_2 - 25 f_4 + 14 f_5)  \frac{e_0^2 c_{F^2 (\nabla
  \phi)^2}}{k_N}\biggr] .  \label{eqn:EMDmass}
\end{multline}
Notice that the Gauss-Bonnet term \emph{does} contribute to the mass correction, unlike in the
4d Reissner-Nordstr{\"o}m case. This is because the dilaton-dependent prefactor
renders it non-topological.

Likewise, using~\eqref{eq.formulaF} we obtain the force correction:
\begin{multline}
  \hat{F}_{\text{self}} = \frac{8 \pi^2}{5} \biggl[ 16(1 + 10 f_2 - 20
  f_3 + 15 f_4 - 6 f_5) c_{\text{GB}} \\ \qquad \qquad +  (1 + 20 f_2 - 80
  f_3 + 75 f_4 - 16 f_5)  \frac{e_0^4 c_{(F^2)^2}}{k_N^2} + 2 (1 - 15 f_4 + 16 f_5)  \frac{e_0^4
  c_{F^4}}{k_N^2} \\
  + 2(1 - 20 f_2 + 75 f_4 - 56 f_5) 
  \frac{e_0^2 c_{F^2 (\nabla \phi)^2}}{k_N} \biggr],  \label{eqn:EMDforce}
\end{multline}
whereas the general entropy result \eqref{eqn:EMDgeneralS} becomes
\begin{equation}
  \delta \mathcal{S}= \frac{8 \pi^2}{\zeta_0}  \biggl( 8 c_{\text{GB}} -
  \frac{e^4_0 c_{F^4}}{k_N^2} \biggr) . \label{eqn:EMDentropy}
\end{equation}
While these complicated functions of $\zeta_0 \equiv \left| \frac{\tilde{e}_0
Q_m}{e_0 Q_e} \right|$ are not particularly interesting in themselves, we note
several important features. First, for $\zeta_0 = 1$ we recover a $\zeta = 1$
dyonic Reissner-Nordstr{\"o}m solution, and in particular \eqref{eqn:EMDmass}
reduces to \eqref{eqn:RNequaldyon} with $c_{R F^2} = 0$.

Second, note that the mass, force and entropy corrections~\eqref{eqn:EMDmass}, \eqref{eqn:EMDforce}, and \eqref{eqn:EMDentropy} each involve linearly-independent combinations of the couplings $c_{\text{GB}}$, $c_{(F^2)^2}$, $c_{F^4}$ and $c_{F^2 (\nabla \phi)^2}$ for every value of $\zeta$, except for the special case $\zeta = 1$ where $\delta \hat{F}_{\text{self}} = - 2 k_N M_0 \delta\! M$ due to the vanishing dilaton charge at the attractor point. Thus, for given charges at a given point in the moduli space, all three corrections are generically independent from each other.

Of course, when viewed as \emph{functions} of the moduli, the mass and force corrections are \emph{not} independent because $\hat{F}_{\text{self}} \equiv e^2_0 Q_e^2 + \tilde{e}^2_0 Q_m^2 - k_N M^2 - 2 \kappa^2_4 \bigl( \frac{d M}{d \phi} \bigr)^2$ depends only on the charges and $M(\phi)$. This implies certain global relations between the signs of the mass and force corrections. For example, suppose there is a unique leading-order attractor point, implying a single continuous family of leading-order extremal solutions as a function of the moduli. In this case, if the force correction is positive (self-repulsive) everywhere in moduli space it follows that the mass correction is negative (super-extremal) everywhere in moduli space, see appendix A of \cite{Harlow:2022gzl}.

To see more explicitly how the mass and force corrections are related in the present example, we substitute $M=M_0 + \delta\!M$ into the definition of $\hat{F}_{\text{self}}$ to obtain
\begin{equation}
  \delta \hat{F}_{\text{self}} = - 2 k_N M_0 \delta\! M - 8 k_N  \frac{d
  M_0}{d \phi}  \frac{d \delta\! M}{d \phi} . \label{eqn:FdMrelation}
\end{equation}
Note that the derivative is taken with respect to the \emph{asymptotic}
value of the modulus, whereas we previously set $\phi_{\infty} = 0$ by
convention. To avoid confusion, it is more convenient to work with $\hat{\phi}
\equiv \phi - \log \zeta_0$, so that the attractor point is fixed at
$\hat{\phi}_h = 0$, whereas $\hat{\phi}_{\infty} = - \log \zeta_0$ is allowed
to vary. Likewise, we re-express the couplings in terms of their fixed,
horizon values
\be
\begin{aligned}
  \hat{c}_{\text{GB}} &= e^{- \phi_h}
  c_{\text{GB}}, & \hat{c}_{(F^2)^2} &= e^{- 3 \phi_h} c_{(F^2)^2}, & \hat{e}_0 &= e^{\phi_h / 2} e_0, \\
  \hat{c}_{F^4} &= e^{- 3 \phi_h} c_{F^4}, & \hat{c}_{F^2 (\nabla \phi)^2}
  &= e^{- 2 \phi_h} c_{F^2 (\nabla \phi)^2},
\end{aligned}
\ee
and we write the leading-order mass $M_0 = \frac{\zeta^{1 / 2}_0 + \zeta_0^{- 1 / 2}}{2}  \hat{M}_0$ in terms of its minimum value
 $\hat{M}_0$ at the attractor point.
In terms of these quantities, \eqref{eqn:EMDmass} becomes
\begin{multline}
  \delta M =  - \frac{4 \pi^2 \zeta^{1 / 2}_0}{5 k_N \hat{M}_0} \biggl[8(2 + 5 f_1 -
  20 f_2 + 20 f_3 - 10 f_4 + 3 f_5) \hat{c}_{\text{GB}}  \\
   \qquad \qquad \qquad  +(1 - 20 f_2 + 40 f_3 -
  25 f_4 + 4 f_5)  \frac{\hat{e}_0^4 \hat{c}_{(F^2)^2}}{k_N^2} + 2(1 + 5 f_4 - 4 f_5) \frac{\hat{e}_0^4 \hat{c}_{F^4}}{k_N^2} \\
   +2(1 - 10 f_1 + 20 f_2 - 25 f_4 + 14 f_5)  \frac{\hat{e}_0^2 \hat{c}_{F^2 (\nabla \phi)^2}}{k_N} \biggr], \label{eqn:EMDmass2}
\end{multline}
where the dependence on $\hat{\phi}_{\infty}$ is now enters entirely through $\zeta_0
= e^{- \hat{\phi}_{\infty}}$. Rewriting~\eqref{eqn:FdMrelation} as
\be
   \delta\hat{F}_{\text{self}} 
  = - 2 k_N \hat{M}_0 \zeta_0  \frac{d}{d \zeta_0}  [(\zeta_0^{1 /
  2} - \zeta_0^{- 1 / 2}) \delta M] ,
\ee
and applying this to~\eqref{eqn:EMDmass2}, one indeed recovers
\eqref{eqn:EMDforce}.\footnote{Note that this equality relies on the absence
of derivative corrections to the relation $\frac{\partial M}{\partial \phi^a}
= - G_{a b}^{\infty} \dot{\phi}^b_{\infty}$, since $\dot{\phi}^a_{\infty}$
(not $\frac{\partial M}{\partial \phi^a}$) was used to derive~\eqref{eq.formulaF}.
While the absence of such corrections was explained in
footnote~\ref{fn:scalarcharge}, we can now see that this is indeed the case in a
non-trivial example.}

Third, note that in the electric limit, $\zeta_0 \rightarrow 0$, the
corrections all diverge:
\be
\begin{aligned}
  \delta M &\rightarrow \frac{2 \pi^2}{15 k_N M}  \biggl( 142 c_{\text{GB}} -
  8 \frac{e_0^4 c_{(F^2)^2}}{k_N^2} - 36 \frac{e_0^4 c_{F^4}}{k_N^2} + 9
  \frac{e_0^2 c_{F^2 (\nabla \phi)^2}}{k_N} \biggr)  \frac{1}{\zeta_0} + O
  (\log \zeta_0),
  \\
  \delta \mathcal{S} &\rightarrow 8 \pi^2 \biggl( 8 c_{\text{GB}} -
  \frac{e^4_0 c_{F^4}}{k_N^2} \biggr)  \frac{1}{\zeta_0} , 
  \\
  \hat{F}_{\text{self}} &\rightarrow 32 \pi^2  \biggl( 8 c_{\text{GB}} -
  \frac{e_0^4 c_{(F^2)^2}}{k_N^2} - 2 \frac{e_0^4 c_{F^4}}{k_N^2} + 2
  \frac{e_0^2 c_{F^2 (\nabla \phi)^2}}{k_N} \biggr) \log (\zeta_0) .
\end{aligned} \label{eqn:elecLimit}
\ee
This is not surprising as the derivative expansion breaks down in this limit,
as previously noted. However, curiously the corrections are all
\emph{finite} in the magnetic limit, $\zeta_0 \rightarrow \infty$:
\be
\begin{aligned}
  \delta M &\rightarrow - \frac{2 \pi^2}{5 k_N M}  \biggl( 16 c_{\text{GB}} +
  \frac{e_0^4 c_{(F^2)^2}}{k_N^2} + 2 \frac{e_0^4 c_{F^4}}{k_N^2} + 2
  \frac{e_0^2 c_{F^2 (\nabla \phi)^2}}{k_N} \biggr), & \delta \mathcal{S} &\rightarrow 0,\\
  \hat{F}_{\text{self}} &\rightarrow \frac{8 \pi^2}{5}  \biggl( 16
  c_{\text{GB}} + \frac{e_0^4 c_{(F^2)^2}}{k_N^2} + 2 \frac{e_0^4
  c_{F^4}}{k_N^2} + 2 \frac{e_0^2 c_{F^2 (\nabla \phi)^2}}{k_N} \biggr) . 
\end{aligned}
\ee
This is because the ``string-frame'' metric $e^{\phi} g_{\mu\nu}$ is non-singular in this limit~\cite{Garfinkle:1990qj,Loges:2019jzs}, taming the string-tree-level derivative corrections. However, since the dilaton blows up near the horizon, ``string loop'' derivative corrections at not similarly tamed, and will give divergent individual contributions, signaling that the derivative expansion does indeed break down near the horizon.

Finally, note that it is possible to choose non-zero couplings $c_{\text{GB}}$, $c_{(F^2)^2}$, $c_{F^4}$, and $c_{F^2 (\nabla \phi)^2}$ such that $\delta M < 0$, $\hat{F}_{\text{self}} > 0$ and $\delta \mathcal{S}> 0$ for arbitrary dyonic charges. For instance, this is the case for the couplings
\begin{equation}
  c_{\text{GB}} = \frac{\alpha'}{16 \kappa^2}, \quad c_{(F^2)^2} =
  \frac{\alpha'}{16} \cdot \frac{5 \kappa^2}{2 e_0^4}, \quad c_{F^4} =
  \frac{\alpha'}{16} \cdot \frac{7 \kappa^2}{4 e_0^4}, \quad c_{F^2 (\nabla
  \phi)^2} = \frac{\alpha'}{16} \cdot \frac{2}{e_0^2},
\end{equation}
given in \S5.5 of \cite{Loges:2019jzs}, where we use $(F \cdot
\tilde{F})^2 = \frac{1}{4} F_{\mu \nu} F^{\nu \rho} F_{\rho \sigma} F^{\sigma
\mu} - \frac{1}{2}  (F^2)^2$ to relate their basis to ours.

\section{Summary and Future Directions} \label{sec.conclusions}

In this paper we obtained new, general formulas for the leading derivative corrections to the mass, entropy and self-force of extremal black holes. We also observed that these corrections are all independent at any given position in the moduli space, complicating earlier attempts to prove that the mass correction is negative by linking it to the entropy correction.

In principle, our results could be used to systematically study the signs of these three corrections in actual quantum gravities, with important implications for various swampland conjectures such as the Weak Gravity Conjecture and the Repulsive Force Conjecture. However, an important obstacle to progress is the fact that relatively little is known about the leading derivative corrections to the low energy effective actions of specific quantum gravities, particularly those in less than ten dimensions. In fact, we are unaware of \emph{any} example where the mass or self-force corrections have been rigorously computed in a specific string theory vacuum to leading non-trivial order in the derivative expansion (the result in~\cite{Kats:2006xp} being questionable due to string loop corrections, see footnote~\ref{fn:dilatonCorrections} and~\cite{Cvetic:1995bj}).

Thus, an extremely interesting (if potentially challenging) direction for future research would be to close the gap between the general effective field theory machinery developed in this paper and actual quantum gravities, or to determine the leading derivative corrections to extremal black holes directly using some more UV-specific tool such as worldsheet techniques. It would also be very interesting to better understand the corrections to extremal black holes whose horizon area vanishes at two-derivative order, though this necessarily requires additional UV input beyond the derivative-corrected low energy effective.

Finally, on a technical level it would be interesting to devise more elegant and efficient derivations of our formulas \eqref{eqn:basicResults}, \eqref{eqn:simpFormulas}, and potentially to generalize them beyond static, spherically symmetric backgrounds. For instance, the ADM formalism~\cite{Arnowitt:1962hi} and/or the Iyer-Wald formalism \cite{Iyer:1994ys,Compere:2018aar,Harlow:2019yfa} (as used in~\cite{Aalsma:2020duv,Aalsma:2021qga}) might provide some of the necessary tools to do so.

\section*{Acknowledgements}
We thank Lars Aalsma, Yue Qiu, Matthew Reece, and Gary Shiu for useful conversations and Matthew Reece and Matteo Lotito for comments on the manuscript. This research was supported by National Science Foundation grants PHY-1914934 and PHY-2112800.

\appendix

\addtocontents{toc}{\protect\setcounter{tocdepth}{1}}

\section{Classifying three- and four-derivative operators\label{sec.basis}}

In this appendix, we classify the possible derivative corrections to the
low-energy effective action~\eqref{eq.twoderivaction} up to four-derivative order.

For the sake of brevity, we only consider parity-invariant operators,\footnote{For our purposes, all moduli $\phi^a$ and gauge fields $A_\mu^A$ have even intrinsic parity.} except
in the four-dimensional case. To justify this omission, note that static,
spherically symmetric electrically charged black holes are parity invariant.
Since mass, self-force, and entropy are parity-even, this implies that
parity-odd operators can only correct these quantities at $O (
{\alpha'}^2 )$. On the other hand, dyonic black holes in four dimensions
\emph{are not} parity invariant, so parity-odd operators \emph{can}
correct their mass, self-force, and entropy at $O (\alpha')$.

Similarly, we do not consider higher-derivative terms with a Lagrangian density that is not gauge and/or general coordinate invariant (e.g., Chern-Simons terms). In particular, such terms typically correspond to topological operators of the form $F \wedge \cdots \wedge F \wedge R \wedge \cdots \wedge R$ in one higher dimension, implying that they are parity-odd and occur only in odd dimensions. If so, they do not contribute by the argument in the previous paragraph.

It is convenient to categorize higher derivative operators by their
``derivative structure'', i.e., the number of first derivatives, second
derivatives, etc., appearing in the operator. Specifically, writing the
operator as $K (\phi)  (\partial^{(n_1)} \phi_1) (\partial^{(n_2)} \phi_2)
\cdots (\partial^{(n_k)} \phi_k)$ for $n_1 \geqslant n_2 \geqslant \cdots
\geqslant n_k > 0$, we abbreviate the derivative structure as $(n_1, \ldots,
n_k)$. Derivative structures can be ordered by ``complexity'', where larger
values of $n_1$ are more complex, with ties broken by the larger value of
$n_2$, further ties broken by the larger value of $n_3$, etc. For instance, by
this classification an operator involving a third derivative is more complex
than one involving any number of second derivatives, whereas an operator
involving multiple second derivatives is more complex than one involving just
one second derivative, and so on.

Since covariant operators often involve a sum of multiple derivative
structures, we label them by their most complex one, e.g., the Ricci scalar
$R$ has derivative structure $(2)$ (even though some terms in it involve only
first derivatives) whereas $F^2$ has derivative structure $(1, 1)$.

After compiling an exhaustive list of operators at a given derivative order,
we can simplify the list in several ways:
\begin{enumerate}
  \item We can impose the Bianchi identities:
  \begin{equation}
    \nabla_{[\mu } F_{ \nu \rho]}^A = 0, \qquad
    \nabla_{[\mu } R^{\nu \rho}_{\;\;\;  \sigma
    \lambda]} = 0, \qquad \nabla_{[\mu } \nabla_{ \nu]}
    \phi^a = 0 .
  \end{equation}
  \item We can replace antisymmetrized covariant derivatives acting on a
  tensor with the Riemann tensor contracted with the tensor.
  
  \item We can integrate by parts.
  
  \item We can impose the leading-order equations of motion:
  \begin{align}
    R_{\mu \nu} & = \kappa^2  \left[ G_{a b} \nabla_{\mu} \phi^a \nabla_{\nu}
    \phi^b + f_{A B} F_{\mu}^a \cdot F_{\nu}^b - \frac{1}{d - 2} g_{\mu \nu}
    f_{A B} F^A \cdot F^B \right], \nonumber\\
    \nabla^{\mu} (f_{A B} F^B_{\mu \nu}) & = 0, \nonumber\\
    \nabla^2 \phi^a & = - \Gamma^a\!_{b c} \nabla \phi^b \cdot
    \nabla \phi^c + \frac{1}{2} G^{a b} f_{A B, b} F^A \cdot F^B . 
  \end{align}
\end{enumerate}
Since the action is not strictly on-shell, the last point requires some
explanation. To be precise, we are free to make field redefinitions involving
derivatives, such as
\begin{equation}
  \phi^a \rightarrow \phi^a + \alpha' \Delta \phi^a,
\end{equation}
where $\Delta \phi^a$ is some operator involving an appropriate number of
derivatives. Then, to first order in $\alpha'$, the action $S = S_2 + \alpha'
\Shd$ changes to
\begin{equation}
  S \rightarrow S_2 + \alpha'  \left( \Shd + \int d^d x \sqrt{- g}
  \Delta \phi^a  \frac{\delta S_2}{\delta \phi^a} \right) + O (
  {\alpha'}^2 ) .
\end{equation}
The leading-order equations of motion are precisely $\frac{\delta S_2}{\delta
\phi^a} = 0$, so in this way we can generate or remove higher-derivative terms
that are proportional to the leading-order equations of motion and/or (after
integration by parts) derivatives of the leading-order equations of motion.

We now proceed as follows. At each derivative order, we first list the
possible operators. Up to four-derivative order, all such operators are built
from the primitive factors
\begin{enumerate}
  \item One derivative: $F$ and $\nabla \phi$,
  
  \item Two derivatives: $R, \nabla F$, and $\nabla^2 \phi$,
  
  \item Three derivatives: $\nabla R, \nabla^2 F$, and $\nabla^3 \phi$,
  
  \item Four derivatives: $\nabla^2 R, \nabla^3 F,$ and $\nabla^4 \phi$,
\end{enumerate}
where we omit Lorentz indices for simplicity for the time being. We then apply
manipulations 1--4 to eliminate more complicated derivative structures in
favor of simpler ones wherever possible. In particular, given any operator
with $n_1 \geqslant n_2 + 2$ we can immediately simplify the derivative
structure via integration by parts, e.g., $(\nabla^2 F) F \rightarrow (\nabla
F)^2$. Up to four-derivative order this eliminates primitive factors
containing more than two derivatives, so we need only consider operators built
from $R, \nabla F, \nabla^2 \phi, F, \nabla \phi$ and arbitrary functions of
the moduli. Moreover, assuming parity, Lorentz indices must be contracted in
pairs, so each operator must contain an even total number of covariant
derivatives $\nabla_{\mu}$ (since $R_{\mu \nu \rho \sigma}$ and $F_{\mu \nu}$
both carry an even number of indices).

\subsection{Parity-even three-derivative operators}

The possible derivative structures at three-derivative order are $(2, 1)$ and
$(1, 1, 1)$. In the former case, we have the possibilities $R F, (\nabla F)
(\nabla \phi)$ and $(\nabla^2 \phi) F$, but only $(\nabla F) (\nabla \phi)$
admits a Lorentz-invariant contraction consistent with the symmetries,
specifically $(\nabla_{\mu} F^{\mu \nu}) (\nabla_{\nu} \phi)$. Since this can
be simplified using the $F$ equations of motion, we can reduce to the $(1, 1,
1)$ derivative structure, where the options are $F^3$ and $F (\nabla \phi)^2$.
Each one admits a unique Lorentz-invariant contraction, hence accounting for
the moduli-dependent prefactors, the complete set of independent parity-even
three derivative operators is
\begin{equation}
  \mathcal{L}_3^{ \text{(even)}} = a_{A B C} (\phi) F^{A \mu}_{\quad \nu} F^{B
  \nu}_{\quad \rho} F^{C \rho}_{\quad \mu} + a_{a b A} (\phi) \nabla^{\mu}
  \phi^a \nabla^{\nu} \phi^b F^A_{\mu \nu} . \label{eqn:L3even}
\end{equation}
$(1, 1, 1)$ is the simplest possible derivative structure at three-derivative
order, hence no further simplifications are possible.

\subsection{Parity-even four-derivative operators} \label{subsec:parityeven}

At four-derivative order, the possible derivative structures are $(2, 2)$,
$(2, 1, 1)$, and $(1, 1, 1, 1)$. We deal with each in turn:

\subsubsection*{Derivative structure $(2, 2)$}

The possibilities are $R^2, R \nabla^2 \phi, (\nabla F) (\nabla F)$, and
$\nabla^2 \phi \nabla^2 \phi$. All but $R^2$ can be simplified, as follows:
\begin{enumerate}
  \item The possible index structures for $R^2$ are
  \begin{equation}
    R^{\mu \nu \rho \sigma} R_{\mu \nu \rho \sigma}, \quad R^{\mu \nu} R_{\mu
    \nu}, \quad \text{or} \quad R^2 .
  \end{equation}
  The latter two can be freely introduced or eliminated using the Einstein
  equations, hence we can transform the first into the Gauss-Bonnet
  combination:
  \begin{equation}
    R_{\text{GB}} \equiv R^{\mu \nu \rho \sigma} R_{\mu \nu \rho \sigma} - 4
    R^{\mu \nu} R_{\mu \nu} + R^2 .
  \end{equation}
  This cannot be further simplified, although it yields a topological operator
  in $d = 4$ (i.e., an operator that is \emph{locally} a total derivative)
  unless multiplied by a moduli-dependent prefactor.
  
  \item Lorentz-invariant contractions of $R \nabla^2 \phi$ always involves
  either the Ricci tensor or the Ricci scalar, so we can transpose them to
  simpler derivative-structures using the Einstein equations.
  
  \item The possible index structures for $(\nabla F) (\nabla F)$ are
  \begin{equation}
    \nabla_{\mu} F^{\mu \nu} \nabla^{\rho} F_{\rho \nu}, \quad \nabla_{\mu}
    F_{\nu \rho} \nabla^{\mu} F^{\nu \rho}, \quad \text{or} \quad \nabla_{\mu}
    F_{\nu \rho} \nabla^{\nu} F^{\rho \mu} .
  \end{equation}
  The first can be simplified using Maxwell's equations whereas the second can
  be transposed into the third using the Bianchi identities and the third can
  be simplified by using integration by parts, commutation of covariant
  derivatives, and Maxwell's equations in turn.
  
  \item In the case of $\nabla^2 \phi \nabla^2 \phi$, the index structure is
  either
  \begin{equation}
    (\nabla_{\mu} \nabla^{\mu} \phi) (\nabla_{\nu} \nabla^{\nu} \phi) \quad
    \text{or} \quad (\nabla_{\mu} \nabla_{\nu} \phi) (\nabla^{\mu}
    \nabla^{\nu} \phi) .
  \end{equation}
  The first can be simplified using the moduli equations of motion, whereas
  the second can be simplified using integrating by parts, commutation of
  covariant derivatives, and the moduli equations of motion.
\end{enumerate}

\subsubsection*{Derivative structure $(2, 1, 1)$}

The possibilities are $R F^2, R (\nabla \phi)^2, (\nabla F) (\nabla \phi) F,
(\nabla^2 \phi) F^2$  and $\nabla^2 \phi (\nabla \phi)^2$. All but $R F^2$ can
be simplified, as follows:
\begin{enumerate}
  \item The possible index structures for $R F^2$ are
  \begin{equation}
    R^{\mu \nu \rho \sigma} F_{\mu \nu} F_{\rho \sigma}, \quad R^{\mu
     \nu} F_{\mu \rho} F_{\nu}^{\;\rho}, \quad \text{or}
    \quad R F_{\mu \nu} F^{\mu \nu} .
  \end{equation}
  The latter two can be simplified using the Einstein equations, whereas the
  first cannot be simplified.
  
  \item Lorentz-invariant index contractions of $R (\nabla \phi)^2$ always
  involves either the Ricci tensor or the Ricci scalar, so we can transpose
  them to simpler operators using the Einstein equations.
  
  \item $(\nabla F) (\nabla \phi) F$ has the possible index structures
  \begin{equation}
    \nabla_{\mu} F^{\mu \nu} \nabla^{\rho} \phi F_{\rho \nu}, \quad
    \nabla_{\mu} F_{\nu \rho} \nabla^{\mu} \phi F^{\nu \rho}, \quad \text{or}
    \quad \nabla_{\mu} F_{\nu \rho} \nabla^{\nu} \phi F^{\mu \rho} .
  \end{equation}
  The first can be simplified using Maxwell's equations, whereas the second
  can be transposed into the third using the Bianchi identities and the third
  can be transposed into an $F^2 \nabla^2 \phi$ term plus a term that can be
  simplified using the Maxwell equations upon integration by parts.
  
  \item $(\nabla^2 \phi) F^2$ has the possible index structures
  \begin{equation}
    (\nabla_{\mu} \nabla^{\mu} \phi) F^{\nu \rho} F_{\nu \rho}, \quad
    \text{or} \quad \nabla_{\mu} \nabla_{\nu} \phi\, F^{\mu \rho}
    F^{\nu}_{\;\; \rho} .
  \end{equation}
  The first can be simplified immediately using the moduli equations of
  motion. To simplify the second, we first integrate by parts, then apply the
  $F$ equations of motion and Bianchi identities, then integrate by parts once
  more:
  \begin{align}
    \nabla_{\mu} \nabla^{\nu} \phi \, F^{\mu \rho}
    F^{\nu}_{\;\; \rho} & \rightarrow - \nabla^{\nu} \phi
    \, (\nabla_{\mu} F^{\mu \rho}) F_{\nu \rho} - \nabla^{\nu}
    \phi \, F^{\mu \rho} \nabla_{\mu} F_{\nu \rho} \nonumber\\
    & \approx - \frac{1}{2} \nabla^{\nu} \phi F^{\mu \rho} \nabla_{\nu}
    F_{\mu \rho} = - \frac{1}{4} \nabla^{\nu} \phi \nabla_{\nu} (F^{\mu \rho}
    F_{\mu \rho}) \rightarrow \frac{1}{4} (\nabla^2 \phi) F^{\mu \rho} F_{\mu
    \rho}, 
  \end{align}
  where ``$\rightarrow$'' means integration by parts and ``$\approx$'' means
  equality up to terms with a simpler derivative structure. The final result
  can now be simplified using the moduli equations of motion.
  
  The above argument assumes that the two gauge fields are the same species.
  More generally,
  \begin{align}
  k_{A B} (\phi) \nabla_{\mu} \nabla^{\nu} \phi \, F^{A \mu
    \rho} F^B_{\nu \rho} &\approx - \frac{1}{2} k_{A B} (\phi) \nabla^{\nu}
    \phi F^{A \mu \rho} \nabla_{\nu} F_{\mu \rho}^B\\
     &= - \frac{1}{4} k_{A B}
    (\phi) \nabla^{\nu} \phi \nabla_{\nu} (F^{A \mu \rho} F_{\mu \rho}^B),
  \end{align}
  where we take $k_{A B} (\phi) = k_{B A} (\phi)$ without loss of generality
  due to the symmetric form of the original operator. Thus, after a further
  integration by parts we can simplify the result as before.
  
  \item $\nabla^2 \phi (\nabla \phi)^2$ has the possible index structures
  \begin{equation}
    (\nabla^{\mu} \nabla_{\mu} \phi) (\nabla^{\nu} \phi) (\nabla_{\nu} \phi),
    \quad \text{or} \quad \nabla^{\mu} \nabla^{\nu} \phi \nabla_{\mu} \phi
    \nabla_{\nu} \phi .
  \end{equation}
  The first can be simplified using the moduli equations. To simplify the
  second, we integrate by parts
  \begin{equation}
    \nabla^{\mu} \nabla^{\nu} \phi \nabla_{\mu} \phi \nabla_{\nu} \phi =
    \frac{1}{2} \nabla^{\mu} (\nabla^{\nu} \phi \nabla_{\nu} \phi)
    \nabla_{\mu} \phi \rightarrow - \frac{1}{2}  (\nabla^{\nu} \phi
    \nabla_{\nu} \phi) (\nabla^{\mu} \nabla_{\mu} \phi),
  \end{equation}
  after which the moduli equations of motion can be used as before. More
  generally, in the presence of multiple moduli:
  \begin{equation}
    k_{a b c} (\phi) \nabla^{\mu} \nabla^{\nu} \phi^a \nabla_{\mu} \phi^b
    \nabla_{\nu} \phi^c \approx k_{a b c} (\phi) \nabla^{\mu} \left[
    \nabla^{\nu} \phi^a \nabla_{\mu} \phi^b \nabla_{\nu} \phi^c - \frac{1}{2}
    \nabla_{\mu} \phi^a \nabla^{\nu} \phi^b \nabla_{\nu} \phi^c \right],
  \end{equation}
  up to terms that can be simplified using the moduli equations of motion,
  where we take $k_{a b c} (\phi) = k_{a c b} (\phi)$ due to the symmetric
  form of the original operator. Thus, after integration by parts we reach the
  simpler $(1, 1, 1, 1)$ derivative structure.
\end{enumerate}

\subsubsection*{Derivative structure $(1, 1, 1, 1)$ and summary}

The possibilities are $F^4, F^2 (\nabla \phi)^2$ and $(\nabla \phi)^4$, with
possible Lorentz-invariant index structures:
\begin{equation}
  (F_{\mu \nu} F^{\mu \nu})^2, \quad F_{\mu \nu} F^{\nu \rho} F_{\rho \sigma}
  F^{\sigma \mu}, \quad F_{\mu \nu} F^{\mu \nu} \nabla_{\rho} \phi
  \nabla^{\rho} \phi, \quad F^{\mu \nu} F_{\mu \rho} \nabla_{\nu} \phi
  \nabla^{\rho} \phi, \quad \text{and} \quad (\nabla_{\mu} \phi \nabla^{\mu}
  \phi)^2 .
\end{equation}
As this is the simplest possible derivative structure at four-derivative
order, none of these can be simplified any further.

Thus, accounting for moduli-dependent prefactors, the complete list of
independent, parity-even four-derivative operators is
\begin{align}
  \mathcal{L}_4^{\text{(even)}} & = a (\phi) R_{\text{GB}} + a_{A B} (\phi)
  R^{\mu \nu \rho \sigma} F_{\mu \nu}^A F_{\rho \sigma}^B + a_{A B C D} (\phi)
  (F^A \cdot F^B) (F^C \cdot F^D) \nonumber \\ &\noeq + b_{A B C D} (\phi) F^A_{\mu \nu} F^{B \nu
  \rho} F^C_{\rho \sigma} F^{D \sigma \mu} 
  + a_{A B a b} (\phi) (F^A \cdot F^B) (\nabla \phi^a \cdot \nabla \phi^b) \nonumber \\ &\noeq
  + b_{A B a b} (\phi) F^{A \mu \nu} F^B_{\mu \rho} \nabla_{\nu} \phi^a
  \nabla^{\rho} \phi^b + a_{a b c d} (\phi) (\nabla \phi^a \cdot \nabla
  \phi^b) (\nabla \phi^c \cdot \nabla \phi^d) . \label{eqn:L4even} 
\end{align}

\subsection{Parity-odd three- and four-derivative operators in $d = 4$}

Parity-odd operators are constructed using the covariant Levi-Civita symbol
$\Omega_{\mu_1 \ldots \mu_d} = \sqrt{- g} \varepsilon_{\mu_1 \ldots \mu_d}$
where $\varepsilon_{\mu_1 \ldots \mu_d} = \pm 1$ is the usual completely
antisymmetric Levi-Civita symbol. Thus the operator must contain $d$
completely antisymmetrized indices. Since at most two indices of $R_{\mu \nu
\rho \sigma}$ can be antisymmetrized, and likewise at most one of the indices
on $\nabla^{(n)} \phi$ can be antisymmetrized (up to terms proportional to the
Riemann tensor), parity-odd operators not involving the gauge fields must have
a derivative order at least as large as the spacetime dimension. In
particular, a complete list of such operators up to 4 derivatives in $d = 4$
is\footnote{Note that both of these operators are topological in the absence
of moduli-dependent prefactors, similar to the 4d Gauss-Bonnet term.}
\begin{equation}
  \mathcal{L}_4^{(\text{odd}, R \phi)} = \tilde{a} (\phi)
  R^{\mu}\!_{ \nu \rho \sigma} R^{\nu}\!_{ \mu \kappa
  \lambda} \Omega^{\rho \sigma \kappa \lambda} + \tilde{a}_{a b c d} (\phi)
  \nabla_{\mu} \phi^a \nabla_{\nu} \phi^b \nabla_{\rho} \phi^c \nabla_{\sigma}
  \phi^d \Omega^{\mu \nu \rho \sigma}, \label{eqn:LoddRphi}
\end{equation}
where there are no such operators at this derivative order for $d > 4$.

For the same reason, once gauge fields are included at least one factor of
$F_{\mu \nu}$ must carry an antisymmetrized index (at the four-derivative
level in $d \geqslant 4$). Consider such an operator
\begin{equation}
  \mathcal{O}=\mathcal{O}_{\mu}^{\nu_2 \ldots \nu_d} F^{\mu \nu_1}
  \Omega_{\nu_1 \nu_2 \ldots \nu_d},
\end{equation}
where $\mathcal{O}_{\mu}^{\nu_2 \ldots \nu_d}$, representing the rest of the
operator, is completely antisymmetric in $\nu_2, \ldots, \nu_d$. Replacing the
indicated factor of $F^{\mu \nu}$ with $\frac{1}{(d - 2) !} \Omega^{\mu \nu
\rho_1 \ldots \rho_{d - 2}}  \tilde{F}_{\rho_1 \ldots \rho_{d - 2}}$ gives
\begin{equation}
  \mathcal{O}= \frac{1}{(d - 2) !} \mathcal{O}_{\mu}^{\nu_2 \ldots \nu_d}
  \Omega^{\mu \nu_1 \rho_1 \ldots \rho_{d - 2}} \Omega_{\nu_1 \nu_2 \ldots
  \nu_d}  \tilde{F}_{\rho_1 \ldots \rho_{d - 2}} = - (d - 1)
  \mathcal{O}_{\mu}^{\mu \rho_1 \ldots \rho_{d - 2}}  \tilde{F}_{\rho_1 \ldots
  \rho_{d - 2}} .
\end{equation}
In this way, we can rewrite the operator in terms of $\tilde{F}_{\mu_1 \ldots
\mu_{d - 2}} = - \frac{1}{2} \Omega_{\mu_1 \ldots \mu_{d - 2} \rho \sigma}
F^{\rho \sigma}$ contracted with other factors, without the explicit
appearance of $\Omega_{\mu_1 \cdots \mu_d}$.

This is particularly convenient in $d = 4$ spacetime dimensions since
$\tilde{F}_{\mu \nu}$ can alternately be viewed as just another species of
gauge field. Thus, reusing our parity-even results, the list of independent
three-derivative parity odd operators in four dimensions is:
\begin{equation}
  \mathcal{L}_{3, d = 4}^{ (\text{odd})} = a_{A B C} (\phi)  \tilde{F}^{A
  \mu}_{\quad \nu} F^{B \mu}_{\quad \nu} F^{C \mu}_{\quad \nu} + a_{a b A}
  (\phi) \nabla^{\mu} \phi^a \nabla^{\nu} \phi^b \tilde{F}^A_{\mu \nu},
\end{equation}
where each term corresponds to a term in (\ref{eqn:L3even}) with a single
factor of $F_{\mu \nu}$ replaced with $\tilde{F}_{\mu \nu}$.

Likewise, at four-derivative order in 4d, the list of parity-odd operators
involving $F_{\mu \nu}$ derived from (\ref{eqn:L4even}) is:
\begin{align}
  \mathcal{L}_{4, d = 4}^{(\text{odd}, F)} & = \tilde{a}_{A B} (\phi) R^{\mu
  \nu \rho \sigma}  \tilde{F}_{\mu \nu}^A F_{\rho \sigma}^B \nonumber\\ &\noeq + \tilde{a}_{A B C
  D} (\phi) (\tilde{F}^A \cdot F^B) (F^C \cdot F^D) + \tilde{b}_{A B C D}
  (\phi) \tilde{F}^A_{\mu \nu} F^{B \nu \rho} F^C_{\rho \sigma} F^{D \sigma
  \mu} \nonumber\\ &\noeq + \tilde{a}_{A B a b} (\phi) (\tilde{F}^A \cdot F^B) (\nabla \phi^a \cdot
  \nabla \phi^b) + \tilde{b}_{A B a b} (\phi) \tilde{F}^{A \mu \nu} F^B_{\mu
  \rho} \nabla_{\nu} \phi^a \nabla^{\rho} \phi^b . \label{eqn:L4oddF} 
\end{align}

In fact, unlike the parity-even case, this list can be reduced still further.
Consider an operator consisting of $(\tilde{F}^A \cdot F^B)$ times another
factor involving at least one index contraction, i.e., of the form
\begin{equation}
  \mathcal{O}= (\tilde{F}^A \cdot F^B) \mathcal{O}_{\lambda}^{\lambda} = -
  \frac{1}{4} \Omega^{\mu \nu \rho \sigma} F^A_{\mu \nu} F^B_{\rho \sigma}
  \mathcal{O}_{\lambda}^{\lambda} .
\end{equation}
Then, since the complete antisymmetrization of 5 indices in $d = 4$ dimensions
vanishes,
\begin{equation}
  0 = 5 \Omega^{[\mu \nu \rho \sigma } F^A_{\mu \nu} F^B_{\rho
  \sigma} \mathcal{O}_{\lambda}^{ \lambda]} = \Omega^{\mu \nu \rho
  \sigma} F_{\mu \nu}^A F_{\rho \sigma}^B \mathcal{O}_{\lambda}^{\lambda} - 2
  \Omega^{\mu \nu \rho \lambda} F^A_{\mu \nu} F_{\rho \sigma}^B
  \mathcal{O}_{\lambda}^{\sigma} - 2 \Omega^{\mu \lambda \rho \sigma} F^A_{\mu
  \nu} F_{\rho \sigma}^B \mathcal{O}_{\lambda}^{\nu},
\end{equation}
and so $\mathcal{O}= \tilde{F}^{A \rho \lambda} F^B_{\rho \sigma}
\mathcal{O}_{\lambda}^{\sigma} + F^A_{\mu \nu} \tilde{F}^{B \mu \lambda}
\mathcal{O}_{\lambda}^{\nu}$.

Two of the operators in (\ref{eqn:L4oddF}) can be eliminated in this way,
leaving the final list of independent, parity-odd four-derivative operators in
4d:
\begin{align}
  \mathcal{L}_{4, d = 4}^{(\text{odd})} & = \tilde{a} (\phi)
  R^{\mu}\!_{\nu \rho \sigma} R^{\nu}\!_{ \mu \kappa
  \lambda} \Omega^{\rho \sigma \kappa \lambda} + \tilde{a}_{A B} (\phi) R^{\mu
  \nu \rho \sigma}  \tilde{F}_{\mu \nu}^A F_{\rho \sigma}^B + \tilde{b}_{A B C
  D} (\phi) \tilde{F}^A_{\mu \nu} F^{B \nu \rho} F^C_{\rho \sigma} F^{D \sigma
  \mu} \nonumber\\
  & \noeq + \tilde{b}_{A B a b} (\phi) \tilde{F}^{A \mu \nu} F^B_{\mu
  \rho} \nabla_{\nu} \phi^a \nabla^{\rho} \phi^b + \tilde{a}_{a b c d} (\phi)
  \nabla_{\mu} \phi^a \nabla_{\nu} \phi^b \nabla_{\rho} \phi^c \nabla_{\sigma}
  \phi^d \Omega^{\mu \nu \rho \sigma}, 
\end{align}
where the first and last entries are from (\ref{eqn:LoddRphi}).

\subsection{Spherically-symmetric backgrounds} \label{subsec:spherical}

In summary, we have found the following three and four-derivative parity-even
operators in general dimension:
\begin{align}
  \mathcal{L}_3^{(\text{even})} & = a_{A B C} (\phi) F^{A \mu}_{\quad \nu}
  F^{B \nu}_{\quad \rho} F^{C \rho}_{\quad \mu} + a_{a b A} (\phi)
  \nabla^{\mu} \phi^a \nabla^{\nu} \phi^b F^A_{\mu \nu}, \nonumber\\
   \mathcal{L}_4^{(\text{even})} & = a (\phi) R_{\text{GB}} + a_{A B} (\phi)
  R^{\mu \nu \rho \sigma} F_{\mu \nu}^A F_{\rho \sigma}^B + a_{A B C D} (\phi)
  (F^A \cdot F^B) (F^C \cdot F^D) \nonumber \\ &\noeq + b_{A B C D} (\phi) F^A_{\mu \nu} F^{B \nu
  \rho} F^C_{\rho \sigma} F^{D \sigma \mu} 
  + a_{A B a b} (\phi) (F^A \cdot F^B) (\nabla \phi^a \cdot \nabla \phi^b) \nonumber \\ &\noeq
  + b_{A B a b} (\phi) F^{A \mu \nu} F^B_{\mu \rho} \nabla_{\nu} \phi^a
  \nabla^{\rho} \phi^b + a_{a b c d} (\phi) (\nabla \phi^a \cdot \nabla
  \phi^b) (\nabla \phi^c \cdot \nabla \phi^d),  \label{eqn:Leven}
\end{align}
as well as the three and four-derivative parity-odd operators in $d = 4$:
\begin{align}
  \mathcal{L}_{3, d = 4}^{ (\text{odd})} & = a_{A B C} (\phi)  \tilde{F}^{A
  \mu}_{\quad \nu} F^{B \mu}_{\quad \nu} F^{C \mu}_{\quad \nu} + a_{a b A}
  (\phi) \nabla^{\mu} \phi^a \nabla^{\nu} \phi^b \tilde{F}^A_{\mu \nu},
  \nonumber\\
  \mathcal{L}_{4, d = 4}^{(\text{odd})} & = \tilde{a} (\phi)
  R^{\mu}\!_{ \nu \rho \sigma} R^{\nu}\!_{\mu \kappa
  \lambda} \Omega^{\rho \sigma \kappa \lambda} + \tilde{a}_{A B} (\phi) R^{\mu
  \nu \rho \sigma}  \tilde{F}_{\mu \nu}^A F_{\rho \sigma}^B + \tilde{b}_{A B C
  D} (\phi) \tilde{F}^A_{\mu \nu} F^{B \nu \rho} F^C_{\rho \sigma} F^{D \sigma
  \mu} \nonumber\\
  & \phantom{=} + \tilde{b}_{A B a b} (\phi) \tilde{F}^{A \mu \nu} F^B_{\mu
  \rho} \nabla_{\nu} \phi^a \nabla^{\rho} \phi^b + \tilde{a}_{a b c d} (\phi)
  \nabla_{\mu} \phi^a \nabla_{\nu} \phi^b \nabla_{\rho} \phi^c \nabla_{\sigma}
  \phi^d \Omega^{\mu \nu \rho \sigma} . 
\end{align}

While these operators are independent in general backgrounds, not all of them
contribute in static, spherically symmetric backgrounds. In particular,
assuming parity, spherical symmetry requires that $F_{t r}^A$ and $\nabla_r
\phi^a$ are the only non-vanishing components of $F_{\mu \nu}^A$ and
$\nabla_{\mu} \phi^a$, respectively, hence evaluating the parity-even
operators (\ref{eqn:Leven}) on a static spherically symmetric background we
obtain:
\begin{equation}
  F^A_{\mu \nu} F^{B \nu \rho} F^C_{\rho \sigma} F^{D \sigma \mu} = 2 (F^A
  \cdot F^B) (F^C \cdot F^D), \quad F^{A \mu \nu} F^B_{\mu \rho} \nabla_{\nu}
  \phi^a \nabla^{\rho} \phi^b = (F^A \cdot F^B) (\nabla \phi^a \cdot \nabla
  \phi^b), \label{eqn:PevenRelns}
\end{equation}
where all the three-derivative operators vanish. In fact, these
relations---which we have observed at the level of the
\emph{action}---persist in the \emph{equations of motion} and in other
first functional derivatives as well. To see why, expand perturbatively in the
spherical-symmetry-breaking components of the various fields,
\begin{equation}
  S = S ^{(0)} + S_{\alpha \beta}^{(2)} \delta \varphi^{\alpha} \delta
  \varphi^{\beta} + \cdots,
\end{equation}
where $\delta \varphi^{\alpha}$ are non-spherically symmetric modes (e.g.,
non-trivial spherical harmonics) and the term linear in $\delta
\varphi^{\alpha}$ is absent due to the underlying spherical symmetry of the
action. Thus, the first functional derivatives of $S$ evaluated on a
spherically symmetric background depend only on $S$ evaluated on a spherically
symmetric background, and the relations implied by spherical symmetric can be
read off from the action itself.

Therefore, up to four-derivative order, the higher-derivative operators making
\emph{independent} $O (\alpha')$ contributions to static, spherically
symmetric, parity-even backgrounds are
\begin{align}
  \mathcal{L}^{\text{(indep)}}_{\leqslant 4} & = a (\phi)
  R_{\text{GB}} + a_{A B} (\phi) R^{\mu \nu \rho \sigma} F_{\mu \nu}^A F_{\rho
  \sigma}^B + a_{A B C D} (\phi) (F^A \cdot F^B) (F^C \cdot F^D) \nonumber\\
  & \phantom{=} + a_{A B a b} (\phi) (F^A \cdot F^B) (\nabla \phi^a \cdot
  \nabla \phi^b) + a_{a b c d} (\phi) (\nabla \phi^a \cdot \nabla \phi^b)
  (\nabla \phi^c \cdot \nabla \phi^d) . \label{eq.Lleq4basis}
\end{align}

In the case of dyonic black holes in $d = 4$, $F_{\mu \nu}^A$ has two
nonvanishing components: $F_{t r}^A$ and $F_{\theta \varphi}^A$. As a result,
while the three-derivative operators still do not contribute, the relations
(\ref{eqn:PevenRelns}) no longer hold. The parity-odd operators are now
relevant as well. However, since the metric and the moduli profiles still
respect parity,\footnote{In particular, dyonic black holes are related to
electric black holes by electromagnetic duality, hence a modified version of
parity is still conserved by dyonic black hole backgrounds where the
modification only involves the gauge fields.} only the parity-odd operators
involving $\tilde{F}$ can contribute. The complete list of higher-derivative
operators making independent $O (\alpha')$ contributions to static,
spherically symmetric dyonic 4d backgrounds is then
\begin{align}
  \mathcal{L}_{\leqslant 4}^{\text{(indep, dyonic)}} & = a (\phi)
  R_{\text{GB}} + a_{A B} (\phi) R^{\mu \nu \rho \sigma} F_{\mu \nu}^A F_{\rho
  \sigma}^B + \tilde{a}_{A B} (\phi) R^{\mu \nu \rho \sigma}  \tilde{F}_{\mu
  \nu}^A F_{\rho \sigma}^B \nonumber\\ &\noeq
   + a_{A B C D} (\phi) (F^A \cdot F^B) (F^C \cdot F^D) + b_{A B
  C D} (\phi) F^A_{\mu \nu} F^{B \nu \rho} F^C_{\rho \sigma} F^{D \sigma \mu} \nonumber\\ &\noeq
  + \tilde{b}_{A B C D} (\phi) \tilde{F}^A_{\mu \nu} F^{B \nu \rho} F^C_{\rho
  \sigma} F^{D \sigma \mu}  + a_{A B a b} (\phi) (F^A \cdot F^B) (\nabla \phi^a \cdot
  \nabla \phi^b) \nonumber\\ &\noeq + b_{A B a b} (\phi) F^{A \mu \nu} F^B_{\mu \rho}
  \nabla_{\nu} \phi^a \nabla^{\rho} \phi^b + \tilde{b}_{A B a b} (\phi)
  \tilde{F}^{A \mu \nu} F^B_{\mu \rho} \nabla_{\nu} \phi^a \nabla^{\rho}
  \phi^b \nonumber\\ &\noeq + a_{a b c d} (\phi) (\nabla \phi^a \cdot \nabla \phi^b)
  (\nabla \phi^c \cdot \nabla \phi^d) . 
\end{align}

\section{Riemann tensor} \label{app:Riemann}
In this appendix, we record the connection coefficients and Riemann tensor for extremal black holes at leading order in $\alpha'$ in our ansatz.
From the extremal metric ansatz 
\begin{equation}
ds^2 =-e^{2\psi(r)} dt^2+e^{-\frac 2{d-3}\psi(r)}\bigl[dr^2+r^2 d\Omega^2_{d-2}\bigr] .
\end{equation}
one obtains the non-vanishing connection coefficients
\begin{align}
\begin{aligned}
\Gamma^t_{tr}&=\psi',\qquad\Gamma^r_{tt}=-g^{rr}g_{tt}\psi',\qquad
\Gamma^r_{rr}=-\frac{\psi'}{d-3}, \\
\Gamma^r_{ij} &=g^{rr}g_{ij} \biggl(\frac{\psi'}{d-3}-\frac{1}{r}\biggr), \qquad
\Gamma^i_{rj} =\biggl(-\frac{\psi'}{d-3}+\frac{1}{r}\biggr) \delta^i_j,\qquad
\Gamma^i_{j k} = \gamma^i_{j k},
\end{aligned}
\end{align}
where $\gamma^i_{j k}$ is the Levi-Civita connection on $S^{d-2}$. One finds the Riemann tensor
\begin{equation}
\begin{aligned}
R^{tr}\!_{ tr}&=-g^{rr}\biggl(\psi''+\frac{d-2}{d-3}(\psi')^2\biggr),&
R^{ti}\!_{ tj}&=g^{rr}\psi'\biggl(\frac{\psi'}{d-3}-\frac{1}{r}\biggr) \delta^i_j,\\
R^{ri}\!_{rj}&=g^{rr}\frac{(r\psi')'}{r(d-3)}\delta^i_j,&
R^{ij}\!_{kl}&=g^{rr}\frac{\psi'}{d-3}\biggl(\frac{2}{r}-\frac{\psi'}{d-3}\biggr) (\delta^i_k \delta^j_l - \delta^i_l \delta^j_k) .
\end{aligned}
\end{equation}
Likewise, the Ricci tensor and Ricci scalar are
\begin{equation}
\begin{aligned}
R^t_t&= -g^{rr}\frac{(r^{d-2}\psi')'}{r^{d-2}}, &
R^r_r&= g^{rr}\biggl(\frac{(r^{d-2}\psi')'}{r^{d-2}(d-3)}-\frac{d-2}{d-3}(\psi')^2\biggr),\\
R^i_j&= 
g^{rr}\frac{(r^{d-2}\psi')'}{(d-3)r^{d-2}} \delta^i_j, & R &= g^{rr}\biggl(\frac{2(r^{d-2}\psi')'}{r^{d-2}(d-3)}-\frac{d-2}{d-3}(\psi')^2\biggr).
\end{aligned}
\end{equation}

\subsection*{Near-horizon limit}

In the near-horizon limit, we have
\begin{equation}
e^\psi = r^{d-3} \biggl(\frac{A_h}{V_{d-2}}\biggr)^{-\frac{d-3}{d-2}} .
\end{equation}
where $A_h$ is the horizon area. The Riemann tensor, Ricci tensor and Ricci scalar then simplify to
\begin{equation}
\begin{aligned}
R^{tr}\!_{ tr}&=R^t_t=R^r_r=-(d-3)^2 \biggl[\frac{V_{d-2}}{A_h}\biggr]^{\frac{2}{d-2}},&
R^{ti}\!_{ tj}&=R^{ri}\!_{rj}= 0,\\
R^{ij}\!_{kl}&=\biggl[\frac{V_{d-2}}{A_h}\biggr]^{\frac{2}{d-2}} (\delta^i_k \delta^j_l - \delta^i_l \delta^j_k), & R^i_j &=(d-3) \biggl[\frac{V_{d-2}}{A_h}\biggr]^{\frac{2}{d-2}} \delta^i_j , \\
R &= -(d-3)(d-4) \biggl[\frac{V_{d-2}}{A_h}\biggr]^{\frac{2}{d-2}} .
\end{aligned}
\end{equation}

\section{Stress tensor, \alt{$\frac{\delta S}{\delta F_{\mu \nu}}$}{dS/dF}, \alt{$\frac{\delta S}{\delta R_{\mu \nu \rho \sigma}}$}{dS/dR}\label{app:stresstensor}}
In this appendix, we record the stress tensor, $\frac{\delta S}{\delta F_{\mu \nu}}$, and $\frac{\delta S}{\delta R_{\mu \nu \rho \sigma}}$ for the higher-derivative action:
\begin{align}
\begin{aligned}
\Shd &=\int d^dx\sqrt{-g} \mathcal L_4,\\
\mathcal L_4&= a R_{\mu\nu\rho\sigma} \hat{R}^{\mu \nu \rho \sigma} 
+a_{AB}R^{\rho \sigma \alpha \beta}F^ A_{\rho \sigma}F^B_{\alpha \beta}
+a_{ABCD}(F^A\cdot F^B)(F^C\cdot F^D) \\
&\noeq+b_{A B C D}F^{A\mu}\!_\nu F^{B\nu}\!_\rho F^{C\rho}\!_\sigma F^{D\sigma}\!_\mu
+a_{ABab}(F^A\cdot F^B)(\nabla \phi^a\cdot \nabla  \phi^b) \\
&\noeq+b_{ABab}F^{A\mu \rho}F^{B\nu}\!_\rho \nabla_\mu\phi^a \nabla_\nu \phi^b + a_{abcd}(\nabla \phi^a\cdot\nabla  \phi^b)( \nabla \phi^c \cdot \nabla \phi^d) ,
\end{aligned}
\label{eq.stressact}
\end{align}
where
\be
\hat{R}^{\mu \nu}_{\;\;\; \rho \sigma} \equiv
  R^{\mu \nu}_{\;\;\; \rho \sigma} - \delta^{\mu}_{\rho} R^{\nu}_{\sigma} +
   \delta^{\mu}_{\sigma} R^{\nu}_{\rho} + \delta^{\nu}_{\rho} R^{\mu}_{\sigma}
   - \delta^{\nu}_{\sigma} R^{\mu}_{\rho} + \frac{1}{2}  (\delta^{\mu}_{\rho}
   \delta^{\nu}_{\sigma} - \delta^{\mu}_{\sigma} \delta^{\nu}_{\rho}) R \,,
\ee
and $R_{\mu\nu\rho\sigma} \hat{R}^{\mu \nu \rho \sigma}=R^{\mu \nu \rho \sigma} R_{\mu \nu \rho \sigma} - 4
    R^{\mu \nu} R_{\mu \nu} + R^2$ is the Gauss-Bonnet density.
Using the notation $\omega_M\stackrel n\circ \xi_N\equiv \omega_M\cdot \xi_N-\frac 1ng_{MN}\omega \cdot \xi$,
one finds
\begin{align}
T_{\mu \nu}&=8 \hat{R}_{\rho \mu \nu \sigma} \nabla^\rho \nabla^\sigma a - 4 a R_{\mu\alpha\beta\gamma} \hat{R}_\nu^{\;\;\alpha\beta\gamma} + g_{\mu \nu} a R_{\alpha\beta\rho\sigma} \hat{R}^{\alpha\beta\rho\sigma} \nonumber \\
&\noeq +4\nabla^\rho\nabla^\sigma\bigl(a_{AB}F^A_{\rho (\mu} F^B_{\nu) \sigma}\bigr)+6 a_{AB}R^{\rho \sigma \alpha }\!_{(\mu}F^A_{\nu)\alpha}F^B_{\rho \sigma} +g_{\mu \nu}a_{AB}R^{\rho \sigma \alpha \beta}F^A_{\rho \sigma}F^B_{\alpha\beta} \nonumber \\
&\noeq -4a_{ABCD}(F^A_\mu \stackrel 4\circ F^B_\nu)(F^C\cdot F^D) \nonumber \\
&\noeq + b_{ABCD}(-8F^A_{\mu \alpha}F^{B\alpha}\!_{\rho}F^{C\rho}\!_\sigma F^{D\sigma}\!_\nu+g_{\mu \nu}F^{A\alpha}\! _\beta F^{B\beta}\!_\rho F^{C\rho}\!_\sigma F^{D\sigma}\!_\alpha) \nonumber \\
&\noeq -2a_{ABab}\biggl[(F^A_\mu \stackrel 4\circ F_\nu^B)(\nabla \phi^a \cdot \nabla \phi^b)+(F^A\cdot F^B)(\nabla_\mu \phi^a \stackrel 4\circ \nabla_\nu \phi^b)\biggr] \nonumber \\
&\noeq -2 b_{A B a b} \biggl[F^A_{\mu \rho} F^B_{\nu \sigma} \nabla^{\rho} \phi^a
  \nabla^{\sigma} \phi^b + 2  F^A_{\rho (\mu } F^{B \rho
  \sigma} \nabla_{ \nu)} \phi^a \nabla_{\sigma} \phi^b
  - \frac{1}{2} g_{\mu \nu}  F^{A}_{\alpha \rho} F_\beta^{B \rho} \nabla^{\alpha} \phi^a \nabla^{\beta} \phi^b\biggr] \nonumber \\
&\noeq -4a_{abcd}(\nabla_\mu \phi^a\stackrel 4\circ \nabla_\nu \phi^b)(\nabla \phi^c\cdot \nabla \phi^d) .
\end{align}
Likewise,
\begin{align}
  \frac{\delta \Shd}{\delta F^A_{\mu \nu}} &= 4 a_{A B} R^{\mu \nu \alpha \beta}
  F_{\alpha \beta}^B + 4 a_{A B C D} F^{B \mu \nu}  (F^C \cdot F^D) + 8 b_{A B
  C D} F^{B \mu \rho} F^C_{\rho \sigma} F^{D \nu \sigma} \nonumber\\
  &\noeq + 2 a_{a b A B} F^{B \mu \nu} (\nabla \phi^a \cdot \nabla \phi^b) - 4
  b_{a b A B} F^{B [\mu }_{\rho} \nabla^{ \nu]} \phi^a
  \nabla^{\rho} \phi^b ,
\end{align}
and
\begin{equation}
    \frac{\delta \Shd}{\delta R_{\mu \nu \rho \sigma}} = 8 a \hat{R}^{\mu \nu \rho \sigma} 
   + \frac{4}{3} a_{A B}  [2 F^{A \mu \nu} F^{B \rho \sigma} \!-\! F^{A \mu
  \rho} F^{B \sigma \nu} \!-\! F^{A \mu \sigma} F^{B \nu \rho}] .
\end{equation}

\bibliographystyle{JHEP}
\bibliography{refs}
\end{document}